\documentclass[11pt]{article}

\usepackage[margin=3cm]{geometry}
\usepackage{caption}
\usepackage{subcaption}
\usepackage{graphicx}%
\usepackage{multirow}%
\usepackage{amsmath,amssymb,amsfonts}%
\usepackage{amsthm}%
\usepackage{mathrsfs}%
\usepackage{latexsym}%
\usepackage{mathtools}%
\usepackage[title]{appendix}%
\usepackage{xcolor}%
\usepackage{color,cancel}
\usepackage{textcomp}%
\usepackage{manyfoot}%
\usepackage[utf8]{inputenc}
\usepackage{listings}%

\usepackage{tocloft}
\usepackage{authblk}
\usepackage{lipsum}
\usepackage{wrapfig}
\usepackage{enumerate}
\usepackage{soul}
\usepackage{chemarrow}

\usepackage{booktabs}%
\usepackage{multirow}
\usepackage[math]{cellspace}
\usepackage{algorithm}%
\usepackage{algorithmicx}%
\usepackage{algpseudocode}%
\usepackage{listings}%
\usepackage{units}
\usepackage{verbatim}
\usepackage{fancyvrb}
\fvset{fontsize=\normalsize}

\usepackage[T1]{fontenc}
\usepackage[sc]{mathpazo}
\setkeys{Gin}{width=\linewidth,totalheight=\textheight,keepaspectratio}

\usepackage{hyperref}
\hypersetup{
	hidelinks = true
}

\newcommand{\lambdags}{\lambda_{\mathrm{g} \to \mathrm{s}}}
\newcommand{\lambdasg}{\lambda_{\mathrm{s} \to \mathrm{g}}}

\newcommand{\micron}{\mu \mathrm{m}}

\usepackage{amsmath}
\usepackage{bm}
\makeatletter
\renewcommand*\env@matrix[1][*\c@MaxMatrixCols c]{%
  \hskip -\arraycolsep
  \let\@ifnextchar\new@ifnextchar
  \array{#1}}
\makeatother

\title{Emergent microtubule properties in a model of filament turnover and nucleation}
    
\author[1]{Anna C Nelson\thanks{\url{annanelson@unm.edu}}}

\author[2]{Scott A McKinley}
\author[3]{Melissa M Rolls}
\author[4,5]{Maria-Veronica Ciocanel}

\affil[1]{Department of Mathematics \& Statistics,  University of New Mexico, Albuquerque, NM, 87131, USA}

\affil[2]{Department of Mathematics, Tulane University, New Orleans, LA, 70118, USA}

\affil[3]{Department of Biochemistry and Molecular Biology, Pennsylvania State University, University Park, PA, 16802, USA}

\affil[4]{Department of Mathematics, Duke University, Durham, NC, 27710, USA}

\affil[5]{Department of Biology, Duke University, Durham, NC, 27710, USA}

\begin{document}
\maketitle
\begin{abstract}
Microtubules (MTs) are dynamic protein filaments essential for intracellular organization and transport, particularly in long-lived cells such as neurons. The plus and minus ends of neuronal MTs switch between growth and shrinking phases, and the nucleation of new filaments is believed to be regulated in both healthy and injury conditions. We propose stochastic and deterministic mathematical models to investigate the impact of filament nucleation and length-regulation mechanisms on emergent properties such as MT lengths and numbers in living cells. We expand our stochastic continuous-time Markov chain model of filament dynamics to incorporate MT nucleation and capture realistic stochastic fluctuations in MT numbers and tubulin availability. We also propose a simplified partial differential equation (PDE) model, which allows for tractable analytical investigation into steady-state MT distributions under different nucleation and length-regulating mechanisms. We find that the stochastic and PDE modeling approaches show good agreement in predicted MT length distributions, and that both MT nucleation and the catastrophe rate of large-length MTs regulate MT length distributions. In both frameworks,  multiple mechanistic combinations achieve the same average MT length. The models proposed can predict parameter regimes where the system is scarce in tubulin, the building block of MTs, and suggest that low filament nucleation regimes are characterized by high variation in MT lengths, while high nucleation regimes drive high variation in MT numbers. These mathematical frameworks have the potential to improve our understanding of MT regulation in both healthy and injured neurons.
\end{abstract}

\begin{footnotesize}
\end{footnotesize}
\section{Introduction}\label{sec:intro}
Microtubules (MTs) are important filament structures within cells that are composed of tubulin dimers and are responsible for intracellular organization and transport. In long-lived and long-range cells like neurons, the MT cytoskeleton is required for sustained transport across large spatial and temporal scales \cite{kelliher2019microtubule,Rolls2021}. Microtubules are polarized structures with chemically-distinct plus and minus ends, and molecular motor proteins are known to carry protein cargo in specified (plus or minus) directions along these filaments. Stable MT structures with specific orientations of their plus and minus ends are critical for sorting and continuous transport of proteins to targeted subcellular destinations. However, the MT cytoskeleton remains dynamic throughout the lifetime of the cell. The lengths of individual MTs grow  and shorten through a process known as dynamic instability, where both the plus and the minus ends stochastically switch between periods of growth and shrinking \cite{mitchison1984dynamic}. Microtubule end dynamics thus regulate the length of the individual filament, which affects the length distribution of a population of filaments. The growth of MTs depends on the availability of tubulin protein dimers. Other microtubule-associated proteins can induce the switch from growth to shrinking, called catastrophe, but also  can stabilize MTs by promoting growth \cite{Rolls2021}. New MTs are nucleated and regulated by $\gamma$-tubulin ring complexes, which act as a template \cite{teixido2012and}. Microtubule nucleation is believed to be tightly controlled throughout the lifetime of living neurons \cite{desai1997microtubule,weiner2021nucleate,luders2021nucleating,vinopal2025centrosomal}.

Due to the complex nature of MT dynamics, mathematical models have long been used to describe MT behavior and give insights into important biological mechanisms. For example, MTs are important for cell division in mammalian cells, where centrosomes position MTs into radial arrays with the minus ends fixed at the nucleation location, forming a centrosomal microtubule organizing center (MTOC). Plus ends of MTs in the MTOC grow and shrink towards sister chromosomes, then attach and aid in chromosomal separation during mitosis. Mathematical studies of this search-and-capture process use mean-field partial differential equation (PDE) models and focus on understanding MT search times in mitosis and cell division \cite{gopalakrishnan2011first,mulder2012microtubules}. Other studies also use PDEs to study the detailed process of MT dynamic instability at their plus ends, where GTP-bound tubulin binds and forms a cap of GTP tubulin; this cap then hydrolyzes and the exposure of GDP tubulin increases the probability of catastrophe events
\cite{Rolls2021,rubin1988,Dogterom1993,zelinski2012dynamics,jemseena2015effects}. Some models also account for MT length regulation mechanisms, such as limited tubulin availability \cite{Dogterom1995,Deymier2005,Margolin2006,hinow2009continuous,Buxton2010,nelson2024minimal} or various length- or age-dependent regulation of MT growth and shrinking \cite{bolterauer1999models,govindan2008length,Mazilu2010,tischer2010providing,kuan2013biophysics,jemseena2015effects,Rank2018}. Fewer studies consider the impact of nucleation on MT length dynamics: in \cite{hinow2009continuous,honore2019growth}, the authors used PDE models to incorporate tubulin-dependent nucleation and growth of MTs, while the authors in \cite{gregoretti2006insights,jain2021polymerization} used stochastic modeling to show that changes in nucleation can impact plus-end MT dynamics and filament lengths. However, understanding emergent MT properties, such as MT length distributions and number of MTs, in complex cell environments requires a rigorous characterization of how multiple mechanisms contribute to these filament properties. In particular, we expect that nucleation, growth and shrinking at both MT ends, and MT length-regulation mechanisms all impact the behavior of non-centrosomal MTs in neuronal cells, where both plus and minus ends are free to shrink and grow. 

While MT dynamics have been studied extensively \textit{in vitro}, here we are interested in understanding MT dynamics and regulation in living neurons. In a previous study, we used stochastic modeling to understand how different length-regulation mechanisms can impact key MT quantities such as MT growth speeds and length distributions \cite{nelson2024minimal}. 
In that study, we were motivated by the experimental evidence that tubulin in MTs turns over every few hours due to dynamic instability \cite{Tao2016}. By analyzing a representative mean-field differential equation model of MT turnover, we parameterized the stochastic model in order to capture realistic \textit{in vivo} behavior. In particular, simulations of this model matched observed distributions of speeds at both MT ends and predicted that MT growth speeds and MT length distributions depend on the mechanisms contributing to length regulation.

Our prior work \cite{nelson2024minimal} assumed a fixed number of MTs, driven by the observation that nucleation is heavily regulated in neurons \cite{chen2012axon,hertzler2020kinetochore}. However, \textit{in vivo} experiments also suggest that nucleation is up-regulated during neuronal injury \cite{Stone2010,chen2012axon,nguyen2014gamma,weiner2020endosomal} and that injury induces a global increase in MT dynamics in both fruit fly and mammalian neurons \cite{chen2012axon,kleele2014assay}. Even in healthy neurons, nucleation can also increase at presynaptic sites by up-stream neuronal activity \cite{qu2019activity}. However, it is not well understood or characterized how various MT growth mechanisms interact with MT nucleation to impact attributes such as the number of MTs and the MT length distributions. Here, we use our model of MT turnover with length-regulating mechanisms at both ends \cite{nelson2024minimal} to study the impact of nucleation on stochastic fluctuations in both tubulin availability and in the MT numbers. We further develop a reduced PDE model that describes the MT dynamics at the plus ends of MTs and allows for variation in MT numbers to analytically determine the relationship between tubulin allocation and average MT lengths and numbers. We show that steady-state analysis of the PDE model provides good agreement with stochastic model results, and that the stochastic model is needed to predict the MT length size--frequency distributions and MT numbers in different nucleation and length-regulation regimes.

\section{Stochastic model of MT growth and nucleation}\label{sec:stochastic}
\begin{figure*}
\includegraphics[width=\textwidth]{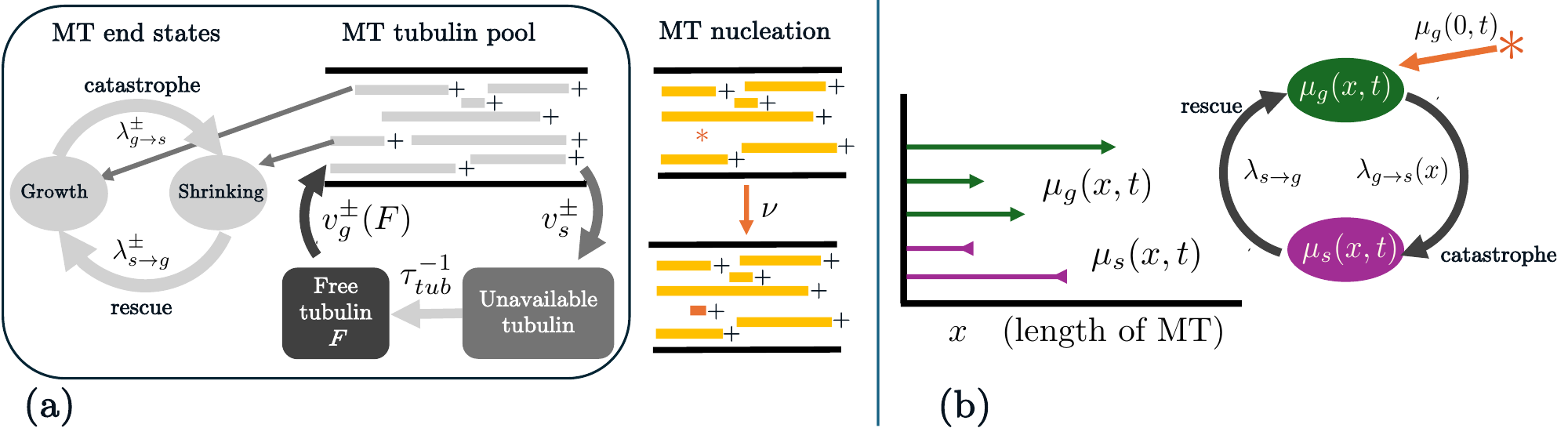}
\caption{(a) Schematic of the CTMC model of MT growth/shrinking dynamics and nucleation, where MT ends stochastically switch between growth and shrinking states and MT growth speeds depend on available free tubulin protein. New MTs appear through the nucleation mechanism, which we model by a Poisson arrival process with arrival rate $\nu$. The grey boxed portion of the schematic was previously developed in \cite{nelson2024minimal} and more details are given in \ref{app:CTMC_model} and \ref{app:algorithm}. (b) Schematic of the PDE model of MT growth dynamics at the plus end. Microtubule plus ends can switch between growth and shrinking phases, where growth depends on tubulin availability and switching from growth to shrinking depends on MT length.  Nucleation of MTs occurs as a boundary condition for the growing MT population of length $0$, i.e. $\mu_{g}(0,t)$.  }
\label{fig:ctmc}
\end{figure*}

To understand how MT nucleation affects MT numbers and turnover dynamics, we first adapt the stochastic continuous-time Markov chain (CTMC) model of MT dynamics we originally developed in \cite{nelson2024minimal}. Figure~\ref{fig:ctmc}(a) shows an outline of this CTMC model framework with nucleation, where the previously-published MT turnover model mechanisms are shown boxed. Our prior MT turnover model assumes a one-dimensional spatial domain that represents a neuronal dendrite segment with uniform polarity. Each MT is described by the pair $(x_+(t),x_-(t))$, where $x_-(t)$ and $x_+(t)$ refer to the position of the minus and plus end, respectively. For all times $t$, we assume $x_+(t)\ge x_-(t)$ and we define the MT length, $x$, as $x = x_+(t) - x_-(t)$. Microtubule ends can be in one of two states: shrinking, where the MT is depolymerizing, or growth, where the MT end is polymerizing. Microtubule ends switch from shrinking to growth at the plus end with rate $\lambda_{s \rightarrow g}^+$ and from growth to shrinking at the plus end with rate $\lambda_{g\rightarrow s}^+$. In the shrinking state, the MT plus ends shrink with velocity $v_s^+$, and in the growth state, the plus ends grow with velocity $v_g^+$.  Similar notation is used to describe the dynamics at the MT minus ends. 

\begin{table}[h]
\begin{centering}
    \centering
    \captionsetup{width = 0.8\textwidth}
    \begin{footnotesize}
    \begin{tabular}{r|c|c| c}
        \hline \hline \textbf{Experimentally-informed quantities} & \textbf{Notation} &{\textbf{Plus-end}}&{\textbf{Minus-end}}\\
        \hline Max growth-phase speed & $v_g^{\mathrm{max}}$ & 9 $\micron/\mathrm{min}$ \cite{Feng2019} &1.125 $\micron/\mathrm{min}$ \cite{Feng2019}\\
        Average growth-phase speed & $\overline{v}_g$ & 6 $\micron/\mathrm{min}$ \cite{Feng2019} &0.75 $\micron/\mathrm{min}$ \cite{Feng2019}\\
                  Average growth phase duration & $\overline{\tau}_g$ & 2 $\mathrm{min}$ \cite{Feng2019}&4 min \cite{Feng2019}\\
        Average growth to shrinking switch & $\lambda_{g \rightarrow s}^0$ & $0.5$ min$^{-1}$ \cite{Feng2019} &0.25 min$^{-1}$  \cite{Feng2019}\\
        Shrinking-phase speed ($|\text{MT}| > 0$) & $\overline{v}_s$ & 6 $\micron/\mathrm{min}$ $\ddagger$ &3.5 $\micron/\mathrm{min}$ $\ddagger$ \\    
        \hline \hline \textbf{Prescribed model parameters} & \textbf{Notation} & {\textbf{Value(s)}} & {\textbf{Source}}\\
        \hline Total available tubulin & $T_\mathrm{tot}$ & {$700 \text{ to } 4000 \, \mu\text{m}$} & this study\\
            MT nucleation rate & $\nu$ & {$0.01\text{ to }10 \text{ min}^{-1}$} &  this study\\
                Length-dependence of catastrophe rate & $\gamma$ & $0 \text{ to } 0.03 \, (\mu\text{m} \cdot \text	{min})^{-1}$ & this study\\
            Target average MT length & $L_*$ &{$35 \mu\text{m}$} &  proposed in \cite{nelson2024minimal} \\
                    Minimum growth to shrinking switch rate & $\lambda_{\text{min}}$ & {$0.1\text{ min}^{-1}$} & estimated from \cite{Feng2019} \\
        Characteristic MT length & $L_0$ &{$35 \mu\text{m}$} & parameterized in \cite{nelson2024minimal}\\
Polymerization/depolymerization speeds &
        $v_g, \, v_s$ &{$v_g^{\max}, \, \overline{v}_s$} & parameterized in \cite{nelson2024minimal} \\
        Catastrophe rates at length $L_0$ &
        $\lambda_{g\to s}^0$ & {$1/\overline{\tau}_{g}$}  & parameterized in \cite{nelson2024minimal} \\
              Michaelis--Menten constant (scarce tubulin) & $F_{1/2}$ &
varies & parameterized in \cite{nelson2024minimal} \\
        Rescue rates &
        $\lambda_{s\to g}$ & varies & parameterized in \cite{nelson2024minimal} 
        \\ \hline
        \end{tabular}
    \caption{Model parameters and experimentally-observed quantities that inform parameter choices. Parameters marked with $\ddagger$ are preliminary estimates from the Rolls Lab. A full description of the parameterization strategy is addressed in~\cite{nelson2024minimal}. \label{tab:nuc_params}}
    \end{footnotesize}
    \end{centering}
\end{table}

In our original model of MT polymerization dynamics, we focused on understanding how two mechanisms contribute to regulating MT length: tubulin-dependent growth and length-dependent catastrophe. These length regulation mechanisms are also relevant for the setting here, where the number of MTs is variable. Figure~\ref{fig:ctmc}(a) illustrates how the two mechanisms are incorporated into our stochastic simulation: the MT growth velocity is assumed to depend on the availability of free tubulin, while the switching rate from growth to shrinking is a function of the length of the MT. Tubulin protein is assumed to be either in MTs ($M(t)$) or available throughout the cell for growth ($F(t)$). Microtubule growth is therefore regulated by the amount of free tubulin and decreases $F(t)$, while shrinking events release tubulin and lead to an increase in $F(t)$. Throughout the simulation, we assume that the amount of tubulin in the cell is fixed, so that $T_{\text{tot}} = M(t) + F(t)$.  Our assumption that total tubulin is fixed is consistent with our prior work \cite{nelson2024minimal}. Tubulin is thought to be constant in the cell and is used to normalize protein levels in other experiments \cite{kshirsagar2024resolving}. However, tubulin levels are difficult to control in experiments and tubulin regulation is not well understood. This assumption allows us to develop a minimal modeling framework and to understand how different tubulin levels impact MT length distributions. 

For tubulin-dependent growth, we model the growth velocity using a Michaelis--Menten mechanism in order to capture the saturating effect of tubulin. Previous mathematical models have used \emph{in vitro} data that show MT growth velocities that depend linearly on concentration, above a critical threshold \cite{walker1988dynamic,white2017exploring}. On the other hand, experimental growth velocities observed \emph{in vivo} have been shown to have a wide distribution \cite{Feng2019}. \emph{In vivo}, the ambient tubulin concentration is unknown. In the absence of precise and fully characterized experimental evidence, our choice of a Michaelis--Menten dependence on available MT length echoes our work in \cite{nelson2024minimal} and is meant to be a phenomenological model for tubulin scarcity. For readers who wish to interpret \emph{in vitro} observations literally, similar conclusions should follow after a conversion to concentration in the style of \cite{white2017exploring}, and an interpretation of $T_\text{tot}$ as the tubulin concentration above the critical threshold. Our choice of Michaelis--Menten form allows for velocities that saturate in excess tubulin environments and that decrease smoothly to a small growth velocity when tubulin is scarce, capturing variation in growth speed with minimal parameter values. Finally, we choose to model the total amount of tubulin in units of length. Since experimental \emph{in vivo} tubulin concentrations are difficult to measure and are currently unknown, we assume that $F_{1/2}, T_{\text{tot}}, M(t),$ and $F(t)$ are measured in microns. However, conversion of tubulin length to concentration is possible as proposed in \cite{white2017exploring}. In the present work, growth velocity on either the plus or minus end of each MT is modeled by

\begin{equation}
\begin{aligned}
  {v_g^\pm( F(t))} & =v_{g}^{\pm,max}\frac{F(t)}{F_{1/2} + F(t)}\\ 
 & = v_{g}^{\pm,max}\frac{T_{\text{tot}} - M(t)}{F_{1/2} + T_{\text{tot}} - M(t)}.\\
    \end{aligned}
\end{equation}

For the length-dependent catastrophe mechanism, we assume that the switching rate from growth to shrinking on either MT end varies linearly with the MT length $x$:
\begin{equation}\label{eq:lengthdepcat}
    \lambdags(x) = \max\left(\lambda_{\min}, \lambdags^0 + \gamma(x-L_0)\right).\,
\end{equation}
Here, $\lambdags^0$ is the experimental growth to shrinking switch rate and $\lambda_{\min}$ is a minimum switch rate that ensures that $\lambda_{g\to s}(x)$ remains positive. The parameter $L_0$ is a characteristic MT length related to the switching rate; if $x = L_0$, the growth to catastrophe switching rate $\lambda_{g\to s}(x)$ is equal to the experimental switching rate, $\lambdags^0$. Meanwhile, $\gamma$ represents the slope of the linear relationship between MT length and switching between growth and catastrophe. Based on previous studies in \cite{nelson2024minimal}, values of $\gamma$ between $0$ and $0.03 \text{ }\mu\text{m}^{-1}\text{min}^{-1}$ resulted in qualitatively different steady-state MT lengths for the same amount of total tubulin, $T_{\text{tot}}$. Therefore, we choose three values of $\gamma$ in that range that represent no length-dependent catastrophe, low length-dependent catastrophe, and high length-dependent catastrophe. Given the expression in equation~\eqref{eq:lengthdepcat}, the switching rate from growth to catastrophe will exceed $\lambda_{\text{min}}$ for all $x$ if $\gamma < (\lambdags^0-\lambda_{\min})/L_0 = \gamma_{\text{crit}}$. Otherwise, the minimum rate $\lambda_{\min}$ will be imposed for short MTs. Since the inverse of the growth-to-catastrophe switch rate is related to the time spent in growth, $\lambda_{\text{min}}$ can be thought of as the longest time spent in growth phase. Experimental data suggests that the longest time spent in growth for the minus end is 10 minutes \cite{Feng2019}. In this study, we choose $\lambda_{\min} = 0.1 \text{ min}^{-1}$, so that $\lambda_{\min} < \lambdags^0$ and $\lambda_{g\to s}(x)$ is nonnegative. 

In the previous stochastic model of MT growth dynamics, we assumed that the number of MTs in the model was fixed. If a MT depolymerized to zero length, we reseeded that MT in the next time step. Here, we model MT nucleation as a Poisson arrival process with arrival rate $\nu$, and if an MT reaches a zero or negative length, we remove that MT from the system. In one time step $\Delta t$, the probability of nucleating $k$ MTs is given by
\begin{equation}
    P({\text{$k$ nucleated MTs in time step $\Delta t$}) = \frac{(\nu \Delta t)^k e^{-\nu \Delta t}}{k!}}\,.
\end{equation}
We assume that each nucleated MT has length $0$, representing the $\gamma$-tubulin ring complex template. In the model, MTs are released from nucleation sites immediately, so we assume that both MT ends start in growth phase after nucleation. Experimental evidence suggests that during nucleation, $\gamma$-tubulin ring complexes cap the minus end and prevent growth, but in neurons, this step seems short-lived as many minus ends in neurons are associated with Patronin$/$CAMSAP proteins. In mammalian neurons, CAMSAP2 recognizes freed minus ends and associates with them as they grow slowly \cite{yau2014microtubule}. We base our model and parameterization on \textit{Drosophila} neurons, where the protein Patronin tracks minus ends as they grow, often for many minutes \cite{Feng2019}. For further details on these mechanisms, as well as our parameterization procedure, see \ref{app:CTMC_model} and \cite{nelson2024minimal}. 

\subsection{Parameterization for the stochastic model}\label{sec:stoch_param}
For the MT turnover model, we use the parameterization procedure outlined in \cite{nelson2024minimal}, which allows us to tune parameters in the stochastic model so as to obtain outputs that match experimental or estimated measurements of MT dynamics at steady state. Table \ref{tab:nuc_params} outlines the experimentally-informed quantities and the prescribed model parameters. We aim to capture known \emph{in vivo} MT behavior, but many parameter values in the stochastic model are unknown due to lack of experimental data. In order to set these unknown parameters so that we recapitulate results seen experimentally, we developed an ODE model representative of the MT growth dynamics seen in the stochastic model. Using a steady-state analysis, we  can obtain relationships between model variables and chose several model parameters that satisfied certain constraints. One key output of this process is a target steady-state MT length, $L_*$ which we do not have experimental data for. Based on \emph{in vivo} cell lengths, we set $L_* = 35\mu$m. For more details on the parameterization procedure and the stochastic simulation, we refer to \cite{nelson2024minimal}.

In the stochastic MT dynamics and nucleation model that we study here, we investigate the key factors that limit MT length (tubulin constraints and length-dependent catastrophe), as well as their interaction with the nucleation mechanism and the variable number of MTs. The number of nucleated MTs is dictated by $\nu$, and while nucleation is thought to be regulated in cells, its experimental value is unknown. In particular \textit{in vivo} experiments show that nucleation can be up- or down-regulated by certain signaling pathways in both healthy and injured neurons \cite{weiner2020endosomal,hertzler2020kinetochore,weiner2021nucleate}. Therefore, we vary $\nu$ by orders of magnitude ($10^{-2}-10^1 \text{ min}^{-1}$) to explore the impact of increasing or decreasing nucleation on MT behavior.  Since a healthy cell could be thought to be at homeostatic equilibrium, we are primarily interested in understanding MT behavior at steady state. The stochastic framework can capture fluctuations in behavior both within individual MT length trajectories and across multiple trajectories. However, this approach requires a high-dimensional parameter space, the exploration of which is computationally intensive and challenging. To overcome these limitations, in section~\ref{sec:pde} we propose a coarse-grained deterministic model, formulated using partial differential equations (PDEs), that captures the dynamics at one end of the MT. 

\section{Reduced PDE model of MT turnover and nucleation}\label{sec:pde}

In the following section, we describe a reduced continuous model that allows for more tractable investigation of MT nucleation and of its interaction with the same mechanisms of length regulation as in the stochastic model described in Section \ref{sec:stochastic}. This reduced modeling framework for MT dynamics consists of PDEs that track densities of growing and shrinking MTs as a function of MT length $x$ and time $t$. These densities provide insights into the size-frequency distributions of MTs lengths over time or at steady state. 

For simplicity, we model MT growth and shrinking at the plus end only in the continuous PDE model framework. Microtubule length $x$ is defined in the semi-infinite domain $x \in [0, \infty)$. Let $\mu_g(x,t)$ and $\mu_s(x,t)$ correspond to densities of growing and shrinking MTs of length $x$ at time $t$. 

The amount of tubulin in MTs at time $t$ is given by:
\begin{equation}\label{eq:total_Mt}
M(t) =  \int_0^\infty x [\mu_g(x,t) + \mu_s(x,t)] dx,
\end{equation}
and $F(t) = T_{\text{tot}} - M(t)$ is the amount of free tubulin. Similarly, the number of MTs at time $t$ is
\begin{equation}\label{eq:total_number}
N(t) =  \int_0^\infty  [\mu_g(x,t) + \mu_s(x,t)] dx \,.
\end{equation}
Polymerizing MTs grow with velocity $v_g$, which depends on the amount of available tubulin, $F(t)$. To model the potential scarcity of available tubulin, we use a Michaelis-Menten formulation so that:
\begin{equation}\label{eq:vgofT}
\begin{aligned}
    v_g( F(t)) &= v_{g}^{max}\frac{F(t)}{F_{1/2} + F(t)} =  v_{g}^{max}\frac{T_{\text{tot}} - M(t)}{F_{1/2} + T_{\text{tot}} - M(t)}\\
    &= v_{g}^{max}\frac{T_{\text{tot}} - \int_0^\infty x [\mu_g(x,t) + \mu_s(x,t)] dx} {F_{1/2} + T_{\text{tot}} - \int_0^\infty x [\mu_g(x,t) + \mu_s(x,t)] dx} \,.\\
    \end{aligned}
\end{equation}
Thus the MT growth velocity is also a function of the amount of tubulin in MTs, i.e. $v_g(M(t))$. 

Both plus and minus ends of MTs undergo periods of growth and shrinkage, with switching between these states. As in Section~\ref{sec:stochastic}, we model this through switching rates between the growth and shrinking states. The rate of switching from shrinking to growing is constant, while the rate of switching from growth to shrinking depends on the MT length $x$, as defined in Eq.~\eqref{eq:lengthdepcat} for the stochastic model.

Together, these mechanisms yield the following time-evolution equations for $\mu_g(x,t)$ and $\mu_s(x,t)$:
\begin{equation} \label{eq:PDE}
\begin{aligned}
    \frac{\partial}{\partial t} {\mu_g}&= - \frac{\partial}{\partial x}\left(v_g(M(t)) \mu_g\right) - \lambda_{g\rightarrow s}(x)\mu_g + \lambda_{s\rightarrow g} \mu_s,\\
     \frac{\partial}{\partial t} {\mu_s}&= \frac{\partial}{\partial x}\left(v_s \mu_s   \right) + \lambda_{g\rightarrow s}(x)\mu_g -\lambda_{s\rightarrow g} \mu_s.
\end{aligned}
\end{equation}
{We remind the reader that, in Eqs.~\eqref{eq:PDE}, $x$ does not denote the location of a MT end, but rather the length of the MT; thus, $\mu_g$ and $\mu_s$ are size-frequency distributions of MT lengths.} The boundary condition for Eqs.~\eqref{eq:PDE} is given by:
\begin{equation}\label{eq:BCs}
\begin{aligned}
    v_g(M(t))\mu_g(0,t) &= \nu,
    \end{aligned}
\end{equation}
which specifies that the rate of nucleation of MTs of length $0$ occurs at rate $\nu$. The $x=0$ boundary is ``non-characteristic'' for the shrinking state population in the sense that the flow of the system hits this boundary with an outward trajectory. (The lengths of the MTs are shrinking to and hit zero in this population if they do not switch to growth phase.) We therefore do not define a value for this boundary and restrict the domain of the PDE to $x \geq 0$ for this reason. This is similar to boundary conditions for models of MT catastrophe in a finite domain considered in \cite{mulder2012microtubules,bressloff2019search,tindemans2010microtubule}.

With the established notation, and in analogy with Eq.~\eqref{eq:total_number}, we can also define the number of growing and shrinking MTs as 
\begin{equation}
\begin{aligned}
    N_g(t) = \int_0^\infty \mu_g(x,t)dx, \qquad
    N_s(t) = \int_0^\infty \mu_s(x,t)dx,
\end{aligned}
\end{equation}
respectively. Then $N(t) = N_g(t) + N_s(t)$ corresponds to the total number of MTs. Similarly, in analogy with Eq.~\eqref{eq:total_Mt}, the amount of tubulin in growing and shrinking MTs is defined to be 
\begin{equation}
\begin{aligned}
    M_g(t) = \int_0^\infty x\mu_g(x,t)dx, \qquad
    M_s(t) = \int_0^\infty x\mu_s(x,t)dx,
\end{aligned}
\end{equation}
respectively. 

\subsection{Steady-state analysis of PDE model}
As with the stochastic model in Section \ref{sec:stoch_param}, we aim to understand how different mechanisms of MT dynamics influence the steady-state solutions of Eqs. \eqref{eq:PDE}. Namely, we wish to solve for $\overline{\mu}_g(x) = \lim_{t \rightarrow \infty} \mu_g(x,t)$ and $\overline{\mu}_s(x) = \lim_{t \rightarrow \infty} \mu_s(x,t)$,  which we refer to as the steady-state MT length distribution of growing and shrinking MTs, respectively. We note that this is not a MT length size probability distribution because it does not integrate to one. Rather, these are frequency distributions, however, for the sake of efficiency, we will not emphasize the word frequency. Then 
\begin{equation}
\begin{aligned}
    \frac{\partial}{\partial t} \overline{\mu}_g &= 0, \qquad
    \frac{\partial}{\partial t} \overline{\mu}_s &= 0,
    \end{aligned}
\end{equation}
so that $\frac{\partial}{\partial t} \overline{\mu} = 0$ and 
\begin{equation}
    \frac{\partial}{\partial t} M(t) = \int_0^\infty x \frac{\partial \overline{\mu}}{\partial t} dx = 0.
\end{equation}
Therefore, the amount of tubulin in MTs is $M(t) \equiv \overline{M}$ and similarly $\overline{T} \equiv T_{\text{tot}} - \overline{M}$. Given Eq.~\eqref{eq:vgofT}, the growth speed at steady state can be written as:
\begin{equation}
\begin{aligned}
    v_g( \overline{M})  &= v_{g}^{max}\frac{T_{\text{tot}} - \overline{M}}{F_{1/2} + T_{\text{tot}} - \overline{M}}\\
    &= v_{g}^{max}\frac{T_{\text{tot}} - \int_0^\infty x[\overline{\mu}_g(x) + \overline{\mu}_s(x)]dx} {F_{1/2} + T_{\text{tot}} - \int_0^\infty x[\overline{\mu}_g(x) + \overline{\mu}_s(x)]dx} \,.
    \end{aligned}
\end{equation}
To find the steady-state solutions $\overline{\mu}_g$ and $\overline{\mu}_s$, we solve:
\begin{equation}\label{eq:PDE_ss}
   \begin{aligned}
  \frac{\partial}{\partial x}\left(v_g(\overline{M}) \overline{\mu}_g\right) & = - \lambda_{g\rightarrow s}(x)\overline{\mu}_g + \lambda_{s\rightarrow g} \overline{\mu}_s,\\
   \frac{\partial}{\partial x}\left(v_s \overline{\mu}_s   \right) &= -\lambda_{g\rightarrow s}(x)\overline{\mu}_g +\lambda_{s\rightarrow g}\overline{\mu}_s.
\end{aligned}
\end{equation}
Subtracting Eqs.~\eqref{eq:PDE_ss} yields 
\begin{equation}
    \frac{\partial}{\partial x} (v_g(\overline{M})\overline{\mu}_g(x) - v_s \overline{\mu}_s(x)) = 0. 
\end{equation}
We then take the definite integral from $x = 0$ to $x = b$ on both sides of the equation, and thus obtain obtain 
\begin{equation}\label{eq:C1}
\left.\left(v_g(\overline{M})\overline{\mu}_g(x) - v_s \overline{\mu}_s(x)\right)\right|_0^b = 0.
\end{equation}
We assume microtubule lengths are bounded; therefore we set 
\begin{equation}\label{eq:bounded_BC}
    \lim_{x\rightarrow \infty} \overline{\mu}_g(x) = \lim_{x\rightarrow \infty} \overline{\mu}_s(x)= 0.
\end{equation}
By taking $b\rightarrow \infty$ in Eq.~\eqref{eq:C1}, we find that 
\begin{equation}
    v_g(\overline{M})\overline{\mu}_g(0) - v_s \overline{\mu}_s(0) = 0,
\end{equation}
which, imposing the $x=0$ boundary condition for $\mu_g$, implies that
\begin{equation}\label{eq:mus_bc}
\overline{\mu}_s(0) = \frac{v_g(\overline{M})}{v_s}\overline{\mu}_g(0) = \frac{\nu}{v_s}.
\end{equation}
From Eq.~\eqref{eq:C1}, the relationship in Eq.~\eqref{eq:mus_bc} holds for all $x$, and thus the MT populations are related through:
\begin{equation}\label{eq:mug_mus}
   \overline{\mu}_s(x) =   \frac{v_g(\overline{M})}{v_s}\overline{\mu}_g(x).
\end{equation}

To solve for the steady-state MT densities, we substitute this relationship into Eq.~\eqref{eq:mug_mus} into Eqs.~\eqref{eq:PDE_ss} and find that $\overline{\mu}_g(x)$ satisfies:
\begin{equation}\label{eq:mu_g}
     \frac{\partial}{\partial x}\left(v_g(\overline{M}) \overline{\mu}_g\right)  = \left(\frac{v_g(\overline{M})\lambdasg}{v_s} - \lambdags(x)\right)\overline{\mu}_g\,,
\end{equation}
hence
\begin{equation}\label{eq:PDE_general}
\begin{aligned}
    \overline{\mu}_g(x) &= c_1 \exp\left(\frac{\lambdasg}{v_s}x - \frac{\int \lambdags(x')dx'}{v_g(\overline{M})}\right),\\
       \overline{\mu}_s(x) &=  c_1 \frac{v_g(\overline{M})}{v_s} \exp\left(\frac{\lambdasg}{v_s}x - \frac{\int \lambdags(x')dx'}{v_g(\overline{M})}\right).
    \end{aligned}
\end{equation}

We note that the expressions for $\overline{\mu}_g(x)$ and $\overline{\mu}_s(x)$ in Eqs.~\eqref{eq:PDE_general} depend on the steady-state amount of tubulin in MTs $\overline{M}$. In turn, $\overline{M}$ also depends on the steady-state solutions for $\overline{\mu}_g(x)$ and $\overline{\mu}_s(x)$:
\begin{equation}\label{eq:Mbar}
\overline{M} =  \int_0^\infty x [\overline{\mu}_g(x) + \overline{\mu}_s(x)]dx \,.
\end{equation}
Equation ~\eqref{eq:Mbar} thus consists of an implicit equation for $\overline{M}$. To solve this, we must find roots $\overline{M}$ that satisfy Eqs.~\eqref{eq:PDE_general} and \eqref{eq:Mbar}. We use the nonlinear root finder \texttt{fzero} in MATLAB and find that a numerical root solution for $\overline{M}$ can be found in the parameter regimes of interest. The steady-state number of MTs and steady-state MT length are then given by:
\begin{equation}\label{eq:Nbar}
\overline{N} =  \int_0^\infty  [\overline{\mu}_g(x) + \overline{\mu}_s(x)]dx \,
\end{equation}
\begin{equation}\label{eq:Lbar}
\overline{L} = \frac{\overline{M}}{\overline{N}}\,,
\end{equation}
respectively.

\subsection{Steady-state length distributions in different length-regulation regimes}
We now derive the solutions for $\overline{\mu}_g(x)$ and $\overline{\mu}_s(x)$ in different regimes of length-dependent catastrophe, as parameterized by $\gamma$ in Eq.~\eqref{eq:lengthdepcat}. Recall from section \ref{sec:stochastic} that we denote $\gamma_{\text{crit}} = \frac{\lambda^0 - \lambda_{\text{min}}}{L_0}$. For $\gamma < \gamma_{\text{crit}}$, the growth to shrinking switching rate is given by $\lambdags(x) = \lambdags^0 + \gamma (x - L_0)$. Using this rate, we solve Eqs.~\eqref{eq:PDE_general} in~\ref{section:ss_length_derivation} and find that the steady-state MT length distributions are:
\begin{equation}\label{eq:PDE_gamma_small}
\begin{aligned}
 \overline{\mu}_g(x) &= 
 \frac{\nu}{v_g(\overline{M})} \exp\left(-\frac{\gamma(x-L_0)^2}{2v_g(\overline{M})} + \left(\frac{\lambdasg}{v_s} - \frac{\lambdags^0}{v_g(\overline{M})}\right)x + \frac{\gamma L_0^2}{2v_g(\overline{M})}\right), & \\
 \overline{\mu}_s(x) &= 
\frac{\nu}{v_s}\exp\left(-\frac{\gamma(x-L_0)^2}{2v_g(\overline{M})}+ \left(\frac{\lambdasg}{v_s} - \frac{\lambdags^0}{v_g(\overline{M})}\right)x + \frac{\gamma L_0^2}{2v_g(\overline{M})}\right). & \\\\
\end{aligned}
\end{equation}
Note that this also holds for $\gamma = 0$, corresponding to no length-dependent catastrophe.

For values of $\gamma > \gamma_{\text{crit}}$, it is possible for the rate of growth to shrinking to be $\lambdags(x) = \lambda_{\text{min}}$. This occurs if the MT length $x$ satisfies
\begin{equation}\label{eq:xcrit}
         x<  L_0 + \frac{\lambda_{\min} - \lambdags^0}{\gamma} = x_{\text{crit}}.
\end{equation}
For biologically-relevant MT lengths, we require $x_{\text{crit}}$ to be nonnegative, which is satisfied for $\gamma > \gamma_{\text{crit}}$. On the other hand, when $x\geq x_{\text{crit}}$, then $\lambdags(x) = \lambdags^0 + \gamma(x-L_0)$. 
Therefore for $\gamma > \gamma_{\text{crit}}$, the steady-state length distributions for growing and shrinking MTs become:
\begin{equation}\label{eq:PDE_gamma_large}
\begin{aligned}
    \overline{\mu}_g(x) &= \begin{cases}
        \frac{\nu}{v_g(\overline{M})} \exp\left(\frac{\lambdasg}{v_s}x - \frac{\lambda_{\text{min}}}{v_g(\overline{M})}x\right), & \text{if } 0 \leq x < x_{\text{crit}},\\
        \frac{\nu}{v_g(\overline{M})}\exp\left(\frac{\lambdasg}{v_s}x - \frac{\lambda_{\min} x_{\text{crit}} + \lambdags^0(x-x_{\text{crit}}) 
    + \frac{\gamma}{2}\left((x-L_0)^2 - (x_{\text{crit}} - L_0)^2\right)}{v_g(\overline{M})}\right), & \text{if } x \geq x_{\text{crit}},
    \end{cases}\\
       \overline{\mu}_s(x) &= \begin{cases}
       \frac{\nu}{v_s}  \exp\left(\frac{\lambdasg}{v_s}x - \frac{\lambda_{\text{min}} }{v_g(\overline{M})}x\right), & \text{if } 0 \leq x < x_{\text{crit}},\\
       \frac{\nu}{v_s} \exp\left(\frac{\lambdasg}{v_s}x - \frac{\lambda_{\min} x_{\text{crit}} + \lambdags^0(x-x_{\text{crit}}) 
    + \frac{\gamma}{2}\left((x-L_0)^2 - (x_{\text{crit}} - L_0)^2\right)}{v_g(\overline{M})}\right), & \text{if } x \geq x_{\text{crit}},
    \end{cases}\\
    \end{aligned}
\end{equation}
respectively. 
~\ref{section:ss_length_derivation} provides derivation details for the final solution for the growing and shrinking MT length  frequency distributions in Eqs.~\eqref{eq:PDE_gamma_small} and \eqref{eq:PDE_gamma_large}, including consideration of the boundedness of MT length size-frequency distributions as imposed in Eq.~\eqref{eq:bounded_BC}.

\begin{figure*}
\includegraphics[width=\textwidth]{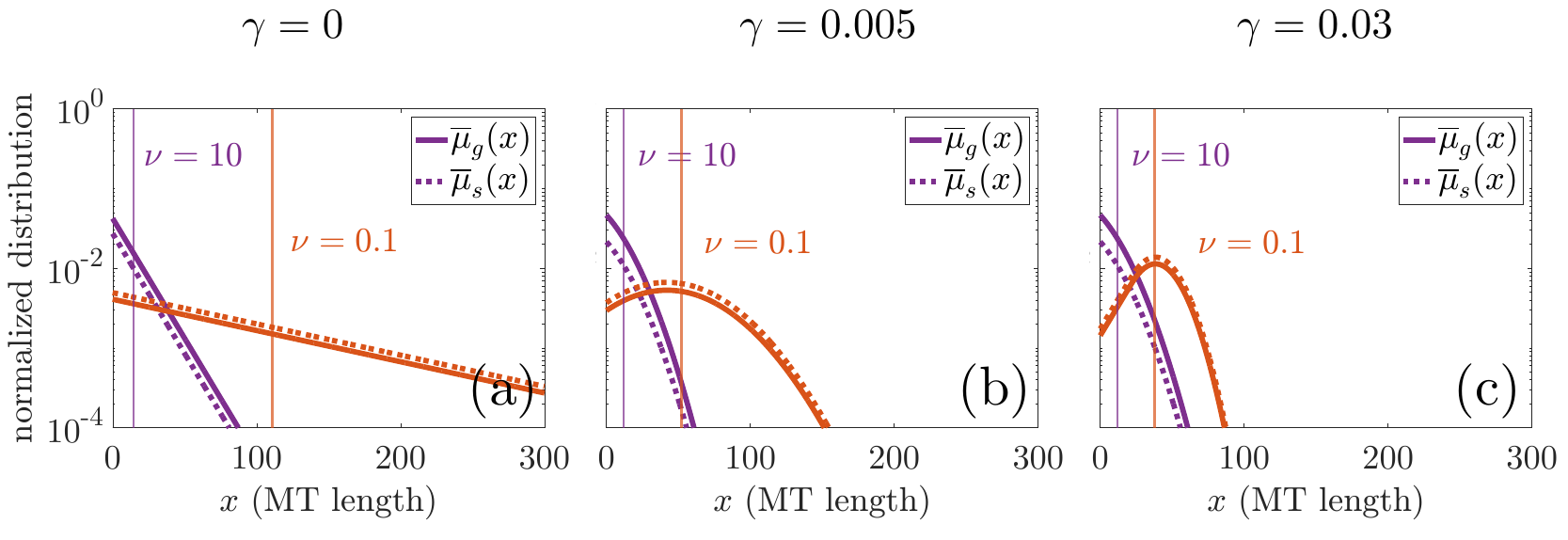}
\caption{PDE model predictions of steady-state normalized length frequency log-distributions of growing (solid line) and shrinking (dashed line) MTs for $T_{\text{tot}} = 1000~\mu$m with (a) no length-dependent catastrophe ($\gamma = 0$), (b) low level of length-dependent catastrophe ($\gamma = 0.005$), and (c) high level of length-dependent catastrophe ($\gamma = 0.03$). Red lines correspond to $\nu = 0.1$ and purple lines correspond to $\nu = 10$. Vertical colored lines indicate the steady-state average MT length, $\overline{L} =  \overline{M}/\overline{N}$, which we will later see is balanced by associated changes in expected MT number.  }
\label{fig:densities}
\end{figure*}

\section{Results}

We study how MT nucleation affects steady-state filament lengths and numbers in the PDE model and the stochastic model. We parameterize both model frameworks to achieve the same target steady-state length ($L_* = 35~\mu$m) with the same baseline total tubulin amount ($T_{\text{tot}} = 1000~\mu$m) \cite{nelson2024minimal}. Additional prescribed model parameters can be found in Table~\ref{tab:nuc_params}. For the PDE model, we solve Eqs.~\eqref{eq:PDE_gamma_small} or \eqref{eq:PDE_gamma_large}, depending on the length-regulating mechanisms considered. Nucleation is implemented by specifying a boundary condition of the PDE system (see Section~\ref{sec:pde}). In the stochastic model, we use the parameterization procedure in \cite{nelson2024minimal} and allow new MTs to be nucleated in the stochastic framework as a Poisson process at a fixed rate (see Section~\ref{sec:stochastic}). The stochastic realizations are run up to 100 hours, and results are collected after one hour of simulation time. We use several values of the nucleation rate and of the length-dependent catastrophe parameter to investigate how these mechanisms impact steady-state MT characteristics. 

The steady-state of the PDE model can be calculated analytically, making it tractable to investigate how nucleation, length-dependent catastrophe, and tubulin limitation impact steady-state MT dynamics, length size-frequency distributions, and overall MT number (Section~\ref{sec:results_PDE}). In addition, solving for the steady-state solutions of this continuous model allows us to more easily study the identifiability of model parameters and to assess if different parameter and mechanistic combinations yield a similar average MT length and MT number.  

The PDE model is easier to analyze; however, it only describes the dynamics at the plus end of the MTs. We are therefore interested in understanding the regimes of parameter space where the simplified PDE model captures similar behaviors as several realizations of the more complicated stochastic model (Section~\ref{sec:results_pde_stoch_lengthdist}). In particular, we compare the resulting MT length distributions, average MT length, and average MT number from both models.

The stochastic model is more complex due to the many parameters and the computational cost of simulating many model realizations to understand how combinations of mechanisms contribute to emerging properties of MT dynamics (Section~\ref{sec:results_stochastic_variance}). However, this framework has the advantage that it can capture realistic fluctuations in MT number and tubulin availability. We seek to understand how nucleation impacts the variance in length and number of MTs given stochastic fluctuations in MT growth dynamics and nucleation. 

\subsection{In the PDE model, small nucleation rates lead to fewer and longer MTs} \label{sec:results_PDE}
We aim to understand how nucleation impacts the predicted steady-state MT length size-frequency distribution in the reduced PDE model in Section~\ref{sec:pde}. Figure~\ref{fig:densities} shows how the normalized densities of steady-state growing MTs, $\overline{\mu}_g(x)/\overline{N}$, and shrinking MTs, $\overline{\mu}_s(x)/\overline{N}$, are impacted by varying the contribution of the length-dependent catastrophe mechanism and by varying nucleation rates. Figure~\ref{fig:densities}a shows that, without length-dependent catastrophe, the size-frequency distribution of steady-state MT lengths for both nucleation values is exponential (note the semi-logarithmic scale.) 

With length-dependent catastrophe, Figure~\ref{fig:densities}b and \ref{fig:densities}c show length size-frequency distributions that are more Gaussian-like, which has also been observed experimentally \cite{gardner2011depolymerizing,gardner2013microtubule, cassimeris1986dynamics,schulze1986microtubule}. All panels in Figure~\ref{fig:densities} show that, as nucleation increases, the steady state average MT length $\overline{L}$ decreases. In the case where length-dependent catastrophe contributes to MT length regulation, the higher nucleation rate leads to a more exponential-like distribution and shorter MT lengths. Interestingly, for the high nucleation rate examined, the size--frequency distributions of MT lengths indicate that there are more growing MTs than shrinking MTs of a given length for all length-dependent catastrophe levels, while the opposite is true for the low nucleation rate. In Eq.~\eqref{eq:PDE_gamma_large}, we note that the growing and shrinking MT length distributions differ only by the constant $\nu/v_g(\overline{M})$ and $\nu/v_s$, respectively. As nucleation increases, our model predicts that the amount of tubulin in MTs, $\overline{M}$, increases. This means that $v_g(\overline{M})$ decreases, and thus explains why the higher nucleation rate results in a larger number of growing MTs at steady-state. For small nucleation values, it is possible for $v_g(\overline{M}) < v_s$, which results in the shrinking MT frequency to be larger than the growing MT frequency, which we see in Figure~\ref{fig:densities}.

\begin{figure*}
\centering
\includegraphics[width=\textwidth]{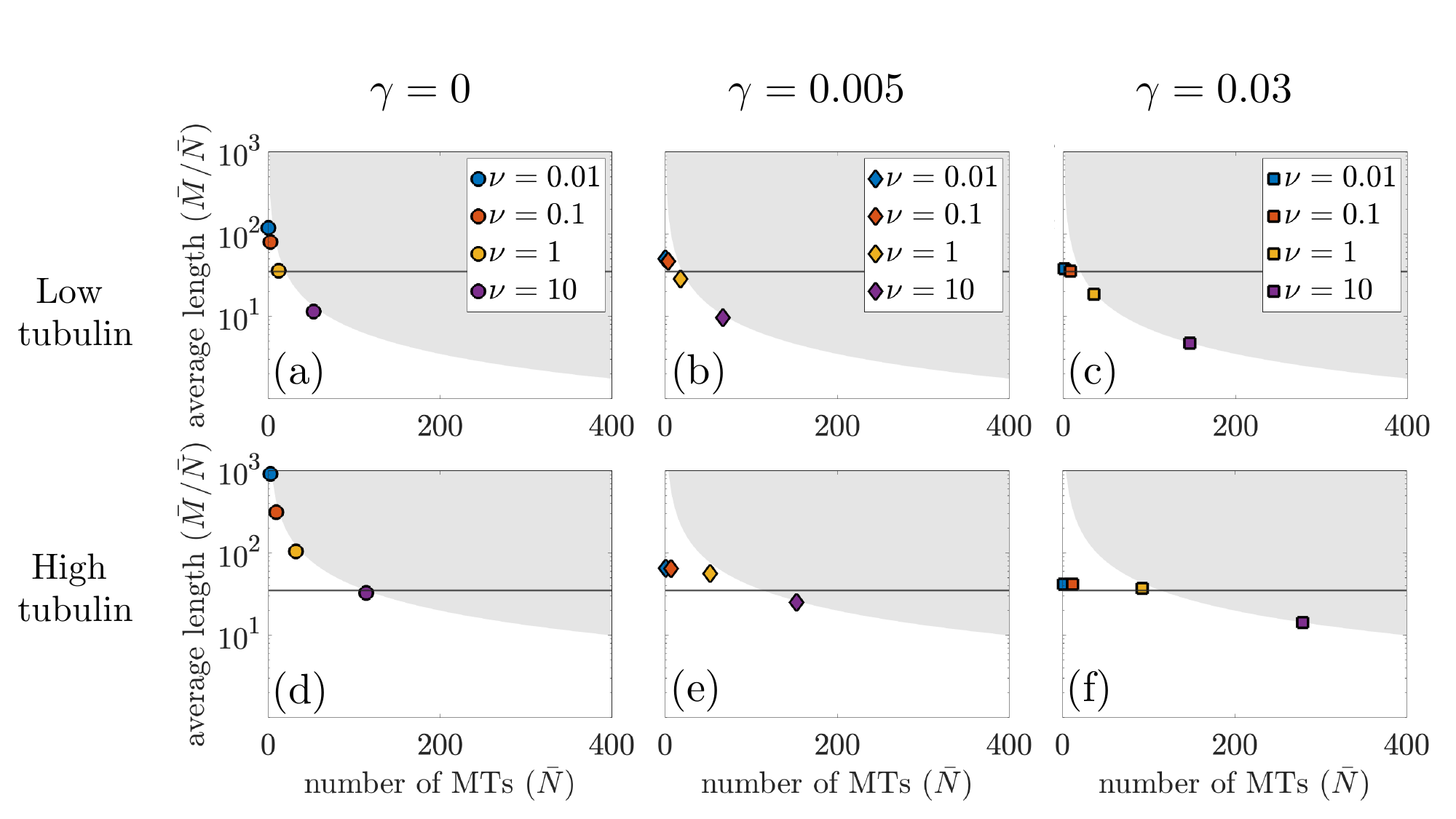}
\caption{Steady-state PDE model results of average number of MTs, $\overline{N}$, versus average length of MTs, $\overline{L}=\overline{M}/\overline{N}$, for $T_{\text{tot}} = 700 \mu$m (top row) and $T_{\text{tot}} = 4000\mu$m (bottom row). Results for varying length-dependent catastrophe are shown in the first, second, and third columns, respectively. The circle, diamond, and square marker correspond to the point $(\overline{N}, \overline{L})$ for $\gamma = 0$, $\gamma = 0.005$, and $\gamma = 0.03$, respectively, defined in Eqs.~\eqref{eq:Nbar} and \eqref{eq:Lbar}.  The grey region corresponds to the infeasible MT lengths and numbers for the given amount of tubulin, where the boundary of this region is found in Eq.~\eqref{eq:tublim}. The horizontal line in each panel represents the target steady-state MT length, $L_*$.}
\label{fig:leng_v_n}
\end{figure*}

Microtubule nucleation thus clearly affects MT length size-frequency distributions, but Eqs.~\eqref{eq:Lbar},~\eqref{eq:Nbar} suggest that they also impact the average number of MTs at steady state, $\overline{N}$, and the average length of MTs at steady state, $\overline{L}$. We also seek to understand how varying the total amount of tubulin available in the cell affects the average number and length of MTs. We therefore investigate how $\overline{N}$ and $\overline{L}$ are affected as $T_{\text{tot}}$ and $\nu$ are varied. As in \cite{nelson2024minimal}, we choose a low tubulin level $T_{\text{tot}} = 700\mu$m and a high tubulin level $T_{\text{tot}} = 4000\mu$m. For the nucleation rate, \textit{in vivo} experiments show that nucleation can vary in both healthy and injured neurons \cite{weiner2020endosomal,hertzler2020kinetochore,weiner2021nucleate}. We therefore vary the nucleation rate $\nu$ across four orders of magnitude ($10^{-2}$ -- $10^{1}$ min$^{-1}$) to understand the impact of this mechanism for a large range of rates. 

Figure~\ref{fig:leng_v_n} shows the average number of MTs versus average steady-state MT length for the chosen tubulin levels and nucleation rates, where each marker denotes the point ($\overline{N},\overline{L}$) for a given level of tubulin and length-dependent catastrophe. For $T_{\text{tot}} = 700\mu$m, this relationship is shown in panels (a,b,c) and for $T_{\text{tot}} = 4000\mu$m, the results are in panels (d,e,f). Each column in Figure~\ref{fig:leng_v_n} represents this relationship for different levels of length-dependent catastrophe. The horizontal line in each panel indicates the target length $L_* = 35\mu$m used to parameterize the models. In particular, the target length impacts the choice of parameters $F_{1/2}, \lambdasg$ in the stochastic and PDE model frameworks. For the steady-state solution of the PDE model, we know that $T_{\text{tot}} = \overline{M} + \overline{T}$, where $\overline{T}$ is the steady-state amount of free tubulin. If all tubulin were contained in MTs, then we would expect that:
\begin{equation}\label{eq:tublim}
    T_{\text{tot}} = \overline{L}\times\overline{N}= \overline{M},
\end{equation}
where $\overline{L}$ is steady-state average MT length. We visualize  this boundary $\overline{L}\times\overline{N}$ for $T_{\text{tot}} = 700\mu$m and $T_{\text{tot}} = 4000\mu$m as a grey region in Figure~\ref{fig:leng_v_n}, representing the infeasible MT lengths and MT numbers for that given amount of tubulin. Note that the closer the markers are to the grey boundary described by Eq.~\eqref{eq:tublim}, the more tubulin-scarce the system is at steady state. 

With no length-dependent catastrophe, Figures~\ref{fig:leng_v_n}(a,d) show that for both low and high tubulin amounts, most tubulin is contained in MTs. For low nucleation rates, the average number of MTs is small, while the average MT length is large. As nucleation increases for both tubulin levels, the average MT length decreases and the average MT number increases, while their relationship tightly follows Eq.~\eqref{eq:tublim}. For each nucleation rate, we see that the average MT length $\overline{L} = \overline{M}/\overline{N}$ is almost an order of magnitude larger for the high tubulin amount compared to the low tubulin amount. Without other length-regulating mechanisms, MT length depends strongly on the total amount of tubulin, as previously seen in \cite{nelson2024minimal}. Given the target average MT length $L_* = 35\mu$m, we find that there exists a nucleation rate $\nu$ that achieves $L_*$ for both tubulin levels. This suggests that multiple combinations of mechanisms can result in the same target MT length. 

With a low amount of length-dependent catastrophe, $\gamma = 0.005 < \gamma_{\text{crit}}$, Figures~\ref{fig:leng_v_n}(b,e) show that average length is 
more tightly regulated compared to the no length-dependent catastrophe case. This is particularly evident in low nucleation cases, where for both $\nu = 0.01 $ and $0.1$, the average lengths for both high and low tubulin do not closely follow the curve from Eq.~\eqref{eq:tublim}. This suggests that the system in this regime is not tubulin scarce and some amount of tubulin is not contained in MTs. Additionally, the average MT length $\overline{L}$ is similar for low levels of nucleation at both tubulin levels. This suggests that the length-dependent catastrophe mechanism coupled with low nucleation leads to a similar average MT length at a variety of tubulin levels. When $\gamma = 0.03>\gamma_{\text{crit}}$, indicating high length-dependent catastrophe, Figures~\ref{fig:leng_v_n}(c,f) show that the average length is even more tightly regulated than the $\gamma = 0.005$ case, with the average MT length close to $L_*$ for almost all nucleation levels. As with the no length-dependent catastrophe case, MT number increases as nucleation level increases for both levels of length-dependent catastrophe.
\begin{figure*}
\includegraphics[width=\textwidth]{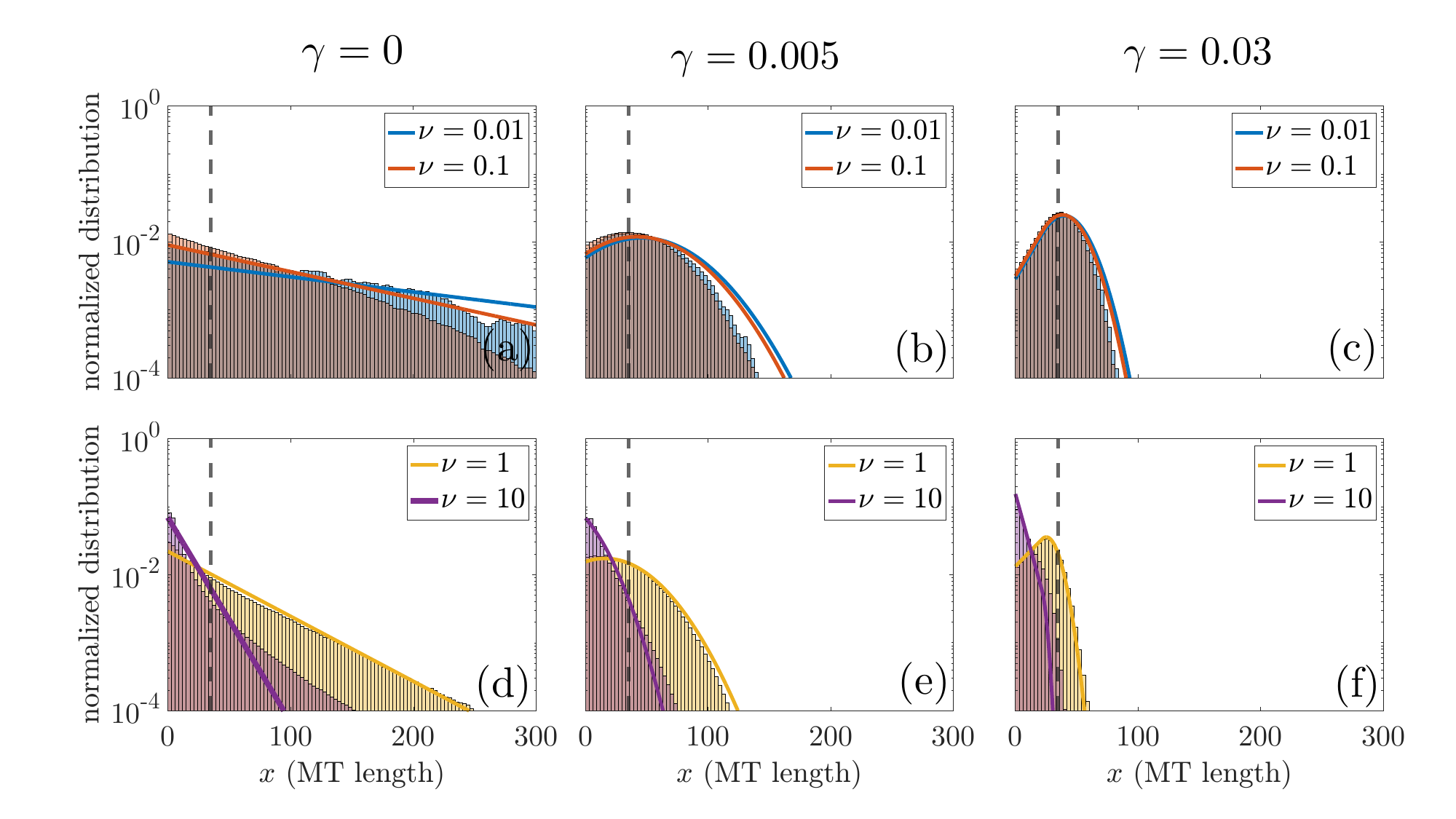}
\caption{Comparison of steady-state MT length log-distributions and empirical stochastic MT length log-distributions for (left) no length-dependent catastrophe, (middle) low length-dependent catastrophe, and (right) high length-dependent catastrophe. Top row shows log-distributions for low nucleation and bottom row illustrates the log-distributions for high nucleation levels. For all cases, $T_{\text{tot}} = 1000\mu$m. Solid lines show steady state densities of $\overline{\mu}_g(x) + \overline{\mu}_s(x)$ for various levels of nucleation corresponding to Eqs.~\eqref{eq:PDE_gamma_small} and \eqref{eq:PDE_gamma_large}, depending on the length-dependent catastrophe mechanism. Histograms are generated from 10 realizations of the MT growth dynamic stochastic model, simulated to 100 hours. }
\label{fig:stoch_vs_ss_distribution}
\end{figure*}

\subsection{Predictions of MT length distributions from the PDE model are consistent with stochastic model simulations}\label{sec:results_pde_stoch_lengthdist}
As discussed in Section~\ref{sec:stochastic}, the stochastic model of MT length regulation and nucleation allows for dynamics on both MT ends and can be computationally expensive to simulate. Given that the reduced PDE model in Section~\ref{sec:results_PDE} is more analytically tractable, we seek to understand whether the solutions of this reduced model capture the key MT dynamics and steady-state properties reflected by the stochastic model. We are also interested in finding whether there are parameter regimes in which the stochastic model predictions deviate from the continuous model predictions. 

In Figure~\ref{fig:stoch_vs_ss_distribution}, we show MT length size-frequency distributions for various nucleation rates obtained from both frameworks. We compare length size-frequency distributions from ten stochastic realizations with the normalized reduced PDE steady-state MT size-frequency distributions ($[\overline{\mu}_g(x) + \overline{\mu}_s(x)]/\overline{N}$) for each nucleation rate from either Eq.~\eqref{eq:PDE_gamma_small} or Eq.~\eqref{eq:PDE_gamma_large}. To compare with the steady-state PDE MT length size-frequency distributions, each realization of the stochastic model is simulated up to 100 hours and we collect MT lengths at every time step (1 second) to construct the stochastic MT length size-frequency distribution.  The dashed vertical line indicates the target MT length of $35\mu$m that was used to set parameter values in both frameworks. We show in the first row the size-frequency distributions for low levels of nucleation and in the second row present results for high levels of nucleation.

For each panel, the shape of the MT length size-frequency distribution for the stochastic simulation agrees well with the results from the reduced PDE framework. The PDE steady-state MT length size-frequency distribution overestimates the frequency of large MT lengths for small nucleation rates, whereas for large nucleation rates, the PDE results underestimate the frequency of large MTs. As in \cite{nelson2024minimal}, the length-dependent catastrophe mechanism influences the type of length distribution at steady state. With no length-dependent catastrophe, Figure~\ref{fig:stoch_vs_ss_distribution}(a,d) shows that for each level of nucleation, the MT length distribution is exponential-like. For low and high levels of length-dependent catastrophe in Figures~\ref{fig:stoch_vs_ss_distribution}(b,c,e,f), both the stochastic and reduced PDE results show a Gaussian-like distribution of MT lengths. As nucleation increases in each panel, we see short MT lengths increase in frequency and the average MT length decreases.
  
\begin{figure*}
\centering
\includegraphics[width=\textwidth]{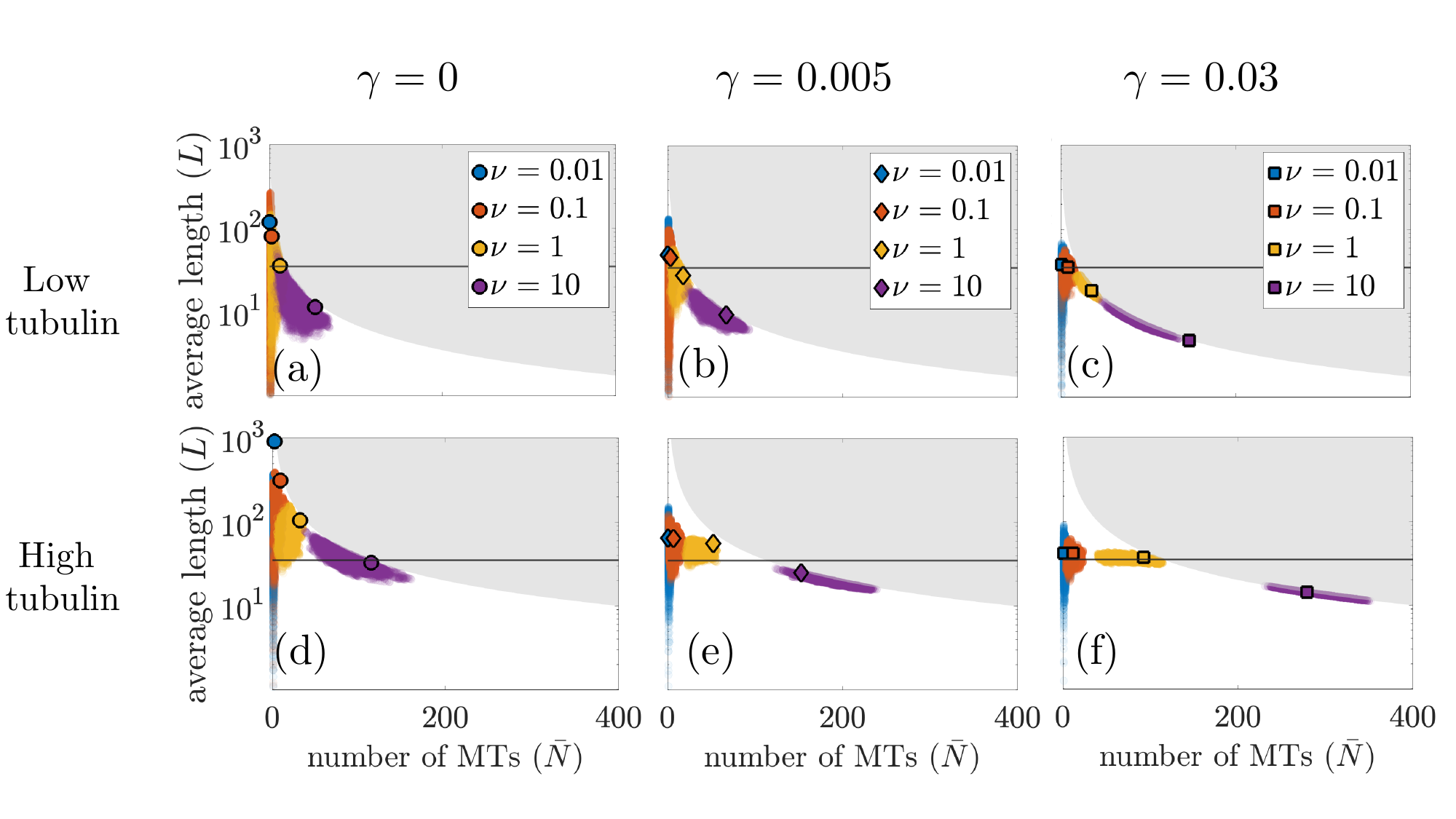}
\caption{Model predictions of the average number of MTs ($\overline{N}$) versus the average length of MTs ($\overline{L}$) from the steady-state PDE solutions (markers outlined in black, same as Fig.~\ref{fig:leng_v_n}) and from 10 realizations of the stochastic model (colored point cloud) simulated up to 5 hours. Each point in the point cloud represents a pair consisting of the MT number and the average MT length $(N,L)$ at each time point (second) of the stochastic simulation. Panels (a,d) show results for $\gamma = 0$, panels (b,e) illustrate results for $\gamma = 0.005$, and panels (c,f) represent results when $\gamma = 0.03$. Results for the low tubulin level ($T_{\text{tot}} = 700\mu$m) are given in the top panels  (a,b,c) and results for the high tubulin level ($T_{\text{tot}} = 4000\mu$m) are found in the bottom panels (d,e,f). The grey area represents infeasible MT lengths and MT numbers for the given tubulin amount, where the border is calculated in Eq.~\eqref{eq:tublim}.}
\label{fig:leng_v_n_stoch}
\end{figure*}

\subsection{The stochastic model predicts high variance in MT length at low nucleation rate and high variance in MT number at high nucleation rate}\label{sec:results_stochastic_variance}

Section~\ref{sec:results_pde_stoch_lengthdist} shows that nucleation impacts the MT length distribution in a similar way in both the reduced PDE model framework and the stochastic framework. From Eqs.~\eqref{eq:Nbar} and~\eqref{eq:Lbar}, we predict that changes in the MT length distribution also influence the steady-state number of MTs and the average MT length. Therefore, we wish to understand how nucleation affects MT number and average MT length with various length-regulating mechanisms. Again, we can compare the results from the stochastic realizations and the reduced PDE model framework to understand the parameter regimes where stochastic fluctuations most impact MT properties such as MT number and length. 

In Figure~\ref{fig:leng_v_n_stoch}, we study how stochastic fluctuations, tubulin availability, and length-dependent catastrophe impact the relationship between average MT length and MT number for both the stochastic and reduced PDE model. The markers outlined in black indicate analytical results from the steady-state solution of the reduced PDE model, while the point clouds represent pairs of MT number and length $(N,L)$ for each time step (1 second) of the stochastic realizations. The top row shows results for low tubulin ($T_{\text{tot}} = 700\mu$m) and the bottom row shows results for high tubulin ($T_{\text{tot}} = 4000\mu$m). In the left column, middle, and right column, we impose the following length-dependent catastrophe parameters for both the PDE and stochastic model: no length-dependent catastrophe ($\gamma = 0$), low length-dependent catastrophe ($\gamma = 0.005$), and high length-dependent catastrophe ($\gamma = 0.03$). As in Figure~\ref{fig:leng_v_n}, the grey region corresponds the region of infeasible MT length and number, given by the tubulin limit in Eq.~\eqref{eq:tublim} for $T_{\text{tot}} = 700 \mu{\text{m}}$ and $T_{\text{tot}} = 4000\mu\text{m}$. 

First, we compare the results from the reduced PDE model with the stochastic model. For nearly all nucleation rates $\nu$, we find that the steady-state PDE results are within their corresponding stochastic simulation point cloud of MT length and number. Additionally, all markers lie below the grey region corresponding to the tubulin limit. When the markers are close to this region, the system is in a tubulin-scarce regime and almost all tubulin is contained in MTs. In the low tubulin regime, Figures~\ref{fig:leng_v_n_stoch}(a,b,c) show that both the stochastic model and the reduced PDE model are fairly close to the tubulin limit. As tubulin increases, the amount of length-dependent catastrophe influences how tubulin-scarce the system is. In Figures~\ref{fig:leng_v_n_stoch}(e,f), both the stochastic and reduced PDE results for lower levels of nucleation are not at the tubulin limit. This indicates that length-dependent catastrophe plays an important role in increasing the amount of available tubulin in the system. However, when nucleation increases to the highest level we studied ($\nu = 10 \text{ min}^{-1}$), the system is again in the tubulin-scarce regime.

While the reduced PDE model can predict a relationship between average MT length and MT number, results from the stochastic simulations illustrate how fluctuations driven by the dynamic instability of MTs impact this relationship. Across tubulin amounts and length-dependent catastrophe levels, there exists a high variance in MT length and low variance in MT number at low nucleation. As the nucleation rate increases, the variance in MT length decreases while the variance in MT number increases. This is particularly evident in the case of high tubulin and high length-dependent catastrophe in Figure~\ref{fig:leng_v_n_stoch}f. For low levels of nucleation, the blue point clouds indicate that MT length can vary almost 2 orders of magnitude in the stochastic model, while the MT number stays relatively fixed. When nucleation is high, the variance in the MT length is relatively small while the MT number varies across time and simulations.

\section{Conclusion and outlook}
In this paper, we propose a stochastic model and a reduced PDE model of MT dynamics and nucleation to understand how MT numbers and length distributions are affected by filament nucleation levels and different length-regulating mechanisms. The stochastic model is parameterized using the procedure developed in~\cite{nelson2024minimal}; here, we also incorporate a MT nucleation mechanism using a Poisson process that allows for new MTs to be seeded. This stochastic framework incorporates fluctuations in both tubulin availability and MT nucleation to understand how these mechanisms impact MT properties such as steady-state length distributions and MT numbers. We compare the results from the stochastic model to steady-state results from a simpler PDE model, which allows for MT nucleation and assumes MT growth dynamics on only one end. 

With two length-regulating mechanisms (tubulin availability and length-dependent catastrophe), our results show that the MT nucleation rate influences both the steady-state MT length distributions and the number of MTs. In most parameter regimes, the MT length distributions from the reduced PDE model agree well with distributions from the stochastic framework. This is the case despite the fact that the PDE model only accounts for MT plus-end dynamics. The similarities in predictions are especially good in regimes with more length-dependent catastrophe regulation. The relationship between average MT length and average MT number is also similar across the stochastic and reduced PDE model frameworks. In particular, as nucleation increases, the average MT length decreases while the average number of MTs increases. 

Since the amount of tubulin is fixed, the inverse relationship between MT length and MT number is to be expected, as there is a limited resource for a variable number of MTs. However, we are also interested in identifying circumstances in which MTs do not utilize all available tubulin. Our results indicate that tubulin scarcity depends on both the strength of the length-dependent catastrophe and on the tubulin amount. As seen in Figure~5, with no length-dependent catastrophe, all nucleation and tubulin levels simulated correspond to a tubulin-scarce result, where most tubulin is contained in MTs. For low tubulin amounts, the system remains tubulin-scarce for all length-dependent catastrophe levels. However, as both length-dependent catastrophe and tubulin amounts increase, we see that there exist nucleation levels where tubulin is not exhausted. Our results indicate that with multiple length-regulating mechanisms, the relationship between MT number and MT length may vary and tubulin allocation may not be straightforward.

As reported before~\cite{nelson2024minimal}, these models of MT dynamics exhibit identifiability issues, since multiple combinations of mechanisms can result in the same target MT length. However, the different mechanistic contributions to filament turnover can lead to different distributions in MT length and number. Our stochastic model results illustrate that, as nucleation increases, the MT length distributions become more exponential-like. In addition, the nucleation rate also impacts the amount of variation observed in MT length and number. For low levels of nucleation, we observe high variation in MT length with a regulated low number of MTs. As nucleation increases, the variability in MT length decreases, while the variation in MT number increases. 

Here, we are particularly motivated by observations of MT dynamics in experiments of living \emph{Drosophila} neurons, where MT growth dynamics and nucleation have been studied in healthy and injured settings \cite{nguyen2014gamma,weiner2021nucleate, Feng2019,Rolls2021}. The results of the models proposed here provide hypotheses that could be validated experimentally. For example, experiments that manipulate tubulin and that up- or down-regulate MT nucleation in neurons have been proposed. The model thus has predictive value for such experiments, as it illustrates that different MT length distributions and distinct regulation of MT numbers can be generated in different mechanistic regimes. In particular, the variance in MT length and number depends on the level of MT nucleation in the cell. Given experimental observations, the model could therefore suggest key potential mechanisms of MT dynamics. 

The distribution of MT lengths has long been of interest in both \textit{in vitro} and \textit{in vivo} experimental settings. \textit{In vitro} studies have reported exponential length distributions \cite{gardner2011depolymerizing,kuo2019spastin}, while \textit{in vivo} and egg extract experiments have resulted in centrosomal MTs with bell-shaped length distributions in control settings \cite{cassimeris1986dynamics,schulze1986microtubule,verde1990regulation,gliksman1992okadaic}. Mathematical modeling studies have sought to understand mechanisms that give rise to observed Gaussian-like distributions of MT lengths \cite{Margolin2006}. Our stochastic and PDE model findings both show that length-dependent catastrophe is sufficient to generate more peaked MT length distributions, confirming our prior results in \cite{nelson2024minimal}. This is consistent with review works that hypothesized that length-dependent depolymerization and MT aging can explain non-exponential lengths of MTs and lead to the tight MT length distributions observed \textit{in vivo} \cite{howard2007microtubule,gardner2013microtubule}. Generally, it is believed that molecular players present in living settings, such as MT (de)polymerases and severing enzymes, regulate MT assembly and are at least partly responsible for peaked MT length distributions \cite{howard2007microtubule,kuo2019spastin}.

More broadly, MTs and other cytoskeleton filaments are polarized structures in healthy neurons (as well as other cells). This polar organization of MT filaments is established and maintained over the course of the lifetime of the organism. Microtubule nucleation is known to be controlled at dendrite branches in neurons and to contribute to the maintenance of polarity. However, it is not known how the stable levels of MT dynamics and organization are maintained over time. Microtuble nucleation is also known to be highly regulated during neural development and altered after neuronal injury. Following axonal injury, MTs are known to display increased nucleation as part of a neuroprotective program \cite{weiner2021nucleate,chen2012axon}. The mathematical models and mechanisms studied here can therefore contribute to our understanding of the mechanisms that control MT polarity and regulation in injury and development.

\section{Acknowledgments}
This work was supported by NIH grant R01NS121245. ACN was partially supported by NSF grant DMS-2038056. ACN would like to thank UNM’s Center for Advanced Research Computing, supported in part by the National Science Foundation, for providing the computing resources and large-scale storage used in this work. 

\appendix
\section{Overview of stochastic model of MT growth dynamics in \cite{nelson2024minimal}}\label{app:CTMC_model}
The stochastic model of MT growth and nucleation discussed in Section~\ref{sec:stochastic} is based on the CTMC model of MT growth dynamics that we previously developed in \cite{nelson2024minimal}. There, we were interested in exploring how two length-limiting mechanisms, length-dependent catastrophe and limited tubulin availability, impact emergent quantities such as MT lengths and distributions of MT speeds for a fixed number of MTs. In particular, we aimed to qualitatively match our stochastic model observables to \emph{in vivo} experimental data such as growth duration and MT growth speed. However, this stochastic framework depends on many parameters related to MT growth and shrinking dynamics, and some of these parameters were more difficult to observe or infer experimentally. Moreover, each realization of the stochastic model was computationally expensive, so we aimed to tune the stochastic model parameters to match available experimental data.

To parameterize the stochastic model, we proposed a continuous model of MT growth and tubulin allocation to understand how model observables depend on parameters for a fixed number of MTs. This ordinary differential equation (ODE) model described the dynamics of the tubulin population (where tubulin can be found in MTs or free), as well as the number of growing and shrinking MTs at both the plus and minus end. We assumed the two length-regulating mechanisms took the same form as described in Section~\ref{sec:stochastic}, and found a unique steady-state solution to the ODE model. This steady-state solution yielded the steady-state MT length, $L_*$, and the steady-state number of growing MTs at the plus and minus end, $G^\pm_*$, all of which depended on prescribed model parameters found in Table~\ref{tab:nuc_params}. 

Our goal was to match outputs of our stochastic model to experimental data using relationships found from the steady-state analysis of the ODE model. In particular, we selected  parameters in our stochastic model so that the following constraints are met: (1) the resulting average MT length in the stochastic framework is equal to the steady-state length, $L_* = \overline{L}$; (2) the average speed of growth phase polymerization is a fraction of the maximum polymerization speeds in a ratio consistent with the observable quantities $\overline{v}_g^\pm$ and $v_g^{\pm,\max}$; and (3) the switching rate from growth to shrinking phase at steady state matches the inverse of the observed growth phase durations $\overline{\tau}_g^{\pm}$. These constraints guided our choice of parameter values in the stochastic model, which resulted in stochastic model outputs that agreed well with experimental data. For example, we set $L_0 = L_*$ so by Eq.~\eqref{eq:lengthdepcat}, $\lambda_{g\rightarrow s}(L_*) = \lambda_{g\rightarrow s}^0 = \frac{1}{\overline{\tau_g}}$ and we satisfy constraint (3).

\section{Algorithm for nucleation and MT dynamics stochastic model implementation}\label{app:algorithm}

To implement the stochastic model of MT growth dyamics and nucleation  in Section~\ref{sec:stochastic}, we utilize Gillespie's $\tau$-leaping scheme \cite{gillespie2001approximate,gillespie2007stochastic}. This scheme assumes that reaction propensities are relatively fixed for a small time step $\Delta t$, namely $[t_i, t_i+\Delta t = t_{i+1}]$. To simulate the model, we first initialize the fixed number of MTs in the domain and all MTs have zero length.  We determine stochastic model parameters given a target average MT length, target MT growth duration, and target average MT growth speed. The parameterization procedure depends on the length-regulation mechanisms implemented. To ensure fixed reaction propensities, we utilize a time step of $1$ second. For every time step, we iterate through the following steps:
\begin{enumerate}
 \item \textbf{Draw growth/shrinking times}: Update the length-dependent catastrophe rate based on the length of the MTs at the previous time step, $t_i$, denoted as $L_i = L(t_i)$. Therefore, the catastrophe rate at $t_{i+1} = t_i + \Delta t$ will take the following form: 
\begin{equation}\label{eq:app_lengthdepcat}
   \lambda^{\pm}_{g\rightarrow s}(L_i) =   \max\left(\lambda_{\min}, \lambdags^\pm + \gamma(L_i - L_0))\right)\,.
\end{equation}
For large $\gamma$ and small $L$ , it is possible that $ \lambda_{\min} > \lambdags^\pm + \gamma(L_i - L_0)$, in whcih case the catastrophe rate will be $\lambda_{\text{min}}$ for short MTs. The updated catastrophe rates are then used to draw times spent in growth phase and times spent in shrinking phase are also drawn. For each end of the MT, the time spent in growth and shrinking, denoted as $t_g$ and $t_s$, respectively, are given by random variables drawn from the appropriate exponential distribution, where
\begin{equation}
    t_{i+1,g} \sim \text{Exp}(\lambda_{g\rightarrow s}(L_i)), \qquad  t_{i+1,s} \sim \text{Exp}(\lambda_{s\rightarrow g}).
    \label{eq:drawtimes}
\end{equation} 
We then determine if the MT exits the time step in either growth or shrinking based on the assumption that only one shrinking phase can occur in a time step $\Delta t$. For example, if a microtubule end entered in growth and $t_g > \Delta t$, then the microtubule would also exit the time step in growth. Similarly, if a MT end entered in growth and $t_g + t_s < \Delta t$, then the MT also exits in growth.

    \item \textbf{Perform shrinking events}: Update the MT lengths and position based on length after shrinking, adding the tubulin to the available tubulin pool, $F$. If the shrinking events resulted in a completely-catastrophed MT (that is, a microtubule with length less than or equal to zero), remove it from the simulation. 
    
    \item \textbf{Perform growth events}: Growth events depend on the available tubulin pool. We thus update the growth length and position based on the available tubulin pool and Michaelis--Menten kinetics. For example, suppose the desired tubulin need for growth for the $j$-th MT is $v_{g} t_g$, where $v_{g}$ is the growth velocity and $t_g$ is the time spent in growth for that MT in time step $\Delta t$. Since growth is dependent on tubulin availability, we update the desired growth length for the $j$-th MT to be
\begin{equation}
    \tilde{x}_j = \frac{F}{F_{1/2} + F}v_g t_g, 
\end{equation}
where $F$ is the available tubulin pool and $F_{1/2}$ is the same Michaelis--Menten constant from Table~\ref{tab:nuc_params}. 

\item \textbf{Perform nucleation events}: Determine the number of nucleation events by drawing from a Poisson distribution with the rate parameter $\lambda = \nu \Delta t$. Nucleated MTs have zero initial length with both ends initialized in growth phase, and are seeded at $x=0$. 
\end{enumerate}

\section{Derivation of steady-state MT length distributions}\label{section:ss_length_derivation}

Based on Eqs.~\eqref{eq:PDE_general}, the steady-state length distributions of growing and shrinking MTs ($\overline{\mu}_g(x)$  and $\overline{\mu}_s(x)$) clearly depend on the length-dependent catastrophe mechanism, which is given by the growth to shrinking rate $\lambda_{g\rightarrow s}(x)$:
\begin{equation}
\lambda_{g\rightarrow s}(x) =  \max\left(\lambda_{\min}, \lambdags^0 + \gamma (x - L_0)\right).
\end{equation}

We first consider the scenario where $\lambda_{g\rightarrow s}(x)$ is always dependent on length. This means that \begin{equation}\label{eq:lambda_gamma_less}
    \lambdags(x) = \lambdags^0 + \gamma(x-L_0),  \quad \forall x \,,
\end{equation}
which in turn means that $\lambda_{\text{min}} < \lambdags^0 + \gamma (x - L_0) $ for all $x$. This occurs when $\gamma \le \frac{\lambdags^0 - \lambda_{\min}}{L_0}= \gamma_{\text{crit}}$. In this case, using the steady-state solutions in Eq.~\eqref{eq:PDE_general} and the switch rate in Eq.~\eqref{eq:lambda_gamma_less} yields the following steady-state length distributions: 
\begin{equation}\label{eq:PDE_general_smallgamma}
\begin{aligned}
    \overline{\mu}_g(x) &= c_1 \exp\left(\frac{\lambdasg}{v_s}x - \frac{\int_0^x \lambdags^0 + \gamma (x' - L_0) dx'}{v_g(\overline{M})}\right),\\
       \overline{\mu}_s(x) &=  c_1 \frac{v_g(\overline{M})}{v_s}
       \exp\left(\frac{\lambdasg}{v_s}x - \frac{\int_0^x \lambdags^0 + \gamma (x' - L_0) dx'}{v_g(\overline{M})}\right)\,.\\
    \end{aligned}
\end{equation}
Evaluating the integrals, we obtain:
\begin{equation}\label{eq:PDE_general_smallgamma_int}
\begin{aligned}
    \overline{\mu}_g(x) 
     &= c_1 \exp\left(\frac{\lambdasg}{v_s}x - \frac{\lambdags x + \frac{\gamma}{2}(x - L_0)^2 - \frac{\gamma L_0^2}{2}}{v_g(\overline{M})}\right),\\
       \overline{\mu}_s(x) 
       &= c_1 \frac{v_g(\overline{M})}{v_s}\exp\left(\frac{\lambdasg}{v_s}x - \frac{\lambdags x + \frac{\gamma}{2}(x - L_0)^2 - \frac{\gamma L_0^2}{2}}{v_g(\overline{M})}\right)\,.\\
    \end{aligned}
\end{equation}
Using the initial condition $v_g(\overline{M})\overline{\mu}_g(0) = \nu$, we find that the constant is given by $c_1 = \frac{\nu}{v_g(\overline{M})}$.

Therefore, for $\gamma \leq \gamma_{\text{crit}}$, the steady-state growing and shrinking MT length distributions are:
\begin{equation}\label{eq:PDE_general_smallgamma_solved}
\begin{aligned}
    \overline{\mu}_g(x) 
     &= \frac{\nu}{v_g(\overline{M})} \exp\left(-\frac{\gamma(x-L_0)^2}{2v_g(\overline{M})} + \left(\frac{\lambdasg}{v_s} - \frac{\lambdags}{v_g(\overline{M})}\right)x + \frac{\gamma L_0^2}{2v_g(\overline{M})}\right),\\
       \overline{\mu}_s(x) 
       &= \frac{\nu}{v_s}\exp\left(-\frac{\gamma(x-L_0)^2}{2v_g(\overline{M})}+ \left(\frac{\lambdasg}{v_s} - \frac{\lambdags}{v_g(\overline{M})}\right)x + \frac{\gamma L_0^2}{2v_g(\overline{M})}\right).\\
    \end{aligned}
\end{equation}

Next, we consider the scenario where $\lambda_{\text{min}} >  \lambdags^0 + \gamma(x-L_0)$, which occurs when
\begin{equation}\label{eq:xcrit}
         x< x_{\text{crit}} = L_0 + \frac{\lambda_{\min} - \lambdags^0}{\gamma}.
\end{equation}
For this regime to be biologically relevant (positive MT lengths), we must have $x_{\text{crit}}>0$. This occurs when $
    \gamma > \gamma_{\text{crit}}=\frac{\lambdags^0 - \lambda_{\min}}{L_0}.$
Therefore, when $\gamma > \gamma_{\text{crit}}$, the growth to shrinking rate is dependent on the MT length in the following way:
\begin{equation}\label{eq:lambda_gamma_great}
    \lambdags(x) = \begin{cases}
        \lambda_{\min} & \text{if } 0 \leq x < x_{\text{crit}},\\
        \lambdags^0 + \gamma(x-L_0) & \text{if } x \geq x_{\text{crit}}\,.
    \end{cases}
\end{equation}
The steady-state growing and shrinking MT length distributions in this setting are:
\begin{equation}\label{eq:PDE_general_largegamma}
\begin{aligned}
    \overline{\mu}_g(x) &= \begin{cases}
        c_1 \exp\left(\frac{\lambdasg}{v_s}x - \frac{\int_0^x \lambda_{\text{min}} dx'}{v_g(\overline{M})}\right), & \text{if } 0 \leq x < x_{\text{crit}},\\
        c_2\exp\left(\frac{\lambdasg}{v_s}x - \frac{\int_0^{x_{\text{crit}}}\lambda_{\min}dx' + \int_{x_{\text{crit}}}^x \lambdags^0 + \gamma(x'-L_0)dx'}{v_g(\overline{M})}\right), & \text{if } x \geq x_{\text{crit}}.
    \end{cases}\\
       \overline{\mu}_s(x) &= \begin{cases}
        c_1\frac{v_g(\overline{M})}{v_s}  \exp\left(\frac{\lambdasg}{v_s}x - \frac{\int_0^x \lambda_{\text{min}} dx'}{v_g(\overline{M})}\right), & \text{if } 0 \leq x < x_{\text{crit}},\\
        c_2\frac{v_g(\overline{M})}{v_s} \exp\left(\frac{\lambdasg}{v_s}x - \frac{\int_0^{x_{\text{crit}}}\lambda_{\min}dx' + \int_{x_{\text{crit}}}^x \lambdags^0 + \gamma(x'-L_0)dx'}{v_g(\overline{M})}\right), & \text{if } x \geq x_{\text{crit}}.
    \end{cases}\\
    \end{aligned}
\end{equation}
Evaluating the integrals yields
\begin{equation}\label{eq:PDE_general_largegamma_int}
\begin{aligned}
    \overline{\mu}_g(x) &= \begin{cases}
        c_1 \exp\left(\frac{\lambdasg}{v_s}x - \frac{\lambda_{\text{min}}}{v_g(\overline{M})}x\right), & \text{if } 0 \leq x < x_{\text{crit}},\\
        c_2\exp\left(\frac{\lambdasg}{v_s}x - \frac{\lambda_{\min} x_{\text{crit}} + \lambdags^0(x-x_{\text{crit}}) 
    + \frac{\gamma}{2}\left((x-L_0)^2 - (x_{\text{crit}} - L_0)^2\right)}{v_g(\overline{M})}\right), & \text{if } x \geq x_{\text{crit}}.
    \end{cases}\\
       \overline{\mu}_s(x) &= \begin{cases}
        c_1\frac{v_g(\overline{M})}{v_s}  \exp\left(\frac{\lambdasg}{v_s}x - \frac{\lambda_{\text{min}} }{v_g(\overline{M})}x\right), & \text{if } 0 \leq x < x_{\text{crit}},\\
        c_2\frac{v_g(\overline{M})}{v_s} \exp\left(\frac{\lambdasg}{v_s}x - \frac{\lambda_{\min} x_{\text{crit}} + \lambdags^0(x-x_{\text{crit}}) 
    + \frac{\gamma}{2}\left((x-L_0)^2 - (x_{\text{crit}} - L_0)^2\right)}{v_g(\overline{M})}\right), & \text{if } x \geq x_{\text{crit}}.
    \end{cases}\\
    \end{aligned}
\end{equation}
Using initial condition $v_g(\overline{M})\overline{\mu}_g(0) = \nu$ and  Eqs.~\eqref{eq:PDE_general_largegamma_int}, we again obtain:
\begin{equation}
     c_1 = \frac{\nu}{v_g(\overline{M})}.
\end{equation}
We require continuity through $\lim_{x \rightarrow x_{\text{crit}}^-}\overline{\mu}_g(x) = \lim_{x \rightarrow x_{\text{crit}}^+}\overline{\mu}_g(x)$, which implies that
\begin{equation}
    \ \frac{\nu}{v_g(\overline{M})}\exp\left(\left(\frac{\lambdasg}{v_s} - \frac{\lambda_{\min}}{v_g(\overline{M})}\right)x_{\text{crit}}\right) = c_2\exp\left(\frac{\lambdasg}{v_s}x_{\text{crit}} - \frac{\lambda_{\min} x_{\text{crit}}  
   }{v_g(\overline{M})}\right), 
\end{equation}
so that
    \begin{equation}
       c_2 = \frac{\nu}{v_g(\overline{M})} = c_1. 
\end{equation}
Therefore, the steady-state solutions in parameter setting are:
\begin{equation}\label{eq:PDE_general_largegamma_solved}
\begin{aligned}
    \overline{\mu}_g(x) &= \begin{cases}
        \frac{\nu}{v_g(\overline{M})} \exp\left(\frac{\lambdasg}{v_s}x - \frac{\lambda_{\text{min}}}{v_g(\overline{M})}x\right), & \text{if } 0 \leq x < x_{\text{crit}},\\
        \frac{\nu}{v_g(\overline{M})}\exp\left(\frac{\lambdasg}{v_s}x - \frac{\lambda_{\min} x_{\text{crit}} + \lambdags^0(x-x_{\text{crit}}) 
    + \frac{\gamma}{2}\left((x-L_0)^2 - (x_{\text{crit}} - L_0)^2\right)}{v_g(\overline{M})}\right), & \text{if } x \geq x_{\text{crit}}.
    \end{cases}\\
       \overline{\mu}_s(x) &= \begin{cases}
       \frac{\nu}{v_s}  \exp\left(\frac{\lambdasg}{v_s}x - \frac{\lambda_{\text{min}} }{v_g(\overline{M})}x\right), & \text{if } 0 \leq x < x_{\text{crit}},\\
       \frac{\nu}{v_s} \exp\left(\frac{\lambdasg}{v_s}x - \frac{\lambda_{\min} x_{\text{crit}} + \lambdags^0(x-x_{\text{crit}}) 
    + \frac{\gamma}{2}\left((x-L_0)^2 - (x_{\text{crit}} - L_0)^2\right)}{v_g(\overline{M})}\right), & \text{if } x \geq x_{\text{crit}}.
    \end{cases}\\
    \end{aligned}
\end{equation}

Note that the solutions in Eqs.~\eqref{eq:PDE_general_smallgamma_solved} and \eqref{eq:PDE_general_largegamma_solved} contain exponential functions of the form $e^{ax^2 + bx + c}$. Since we seek bounded steady-state MT lengths, we note that:
\begin{equation}
    \lim_{x\rightarrow\infty} e^{ax^2 + bx + c} = \begin{cases}
\infty, & \text{if } a >0\\
0,& \text{if }  a < 0.
\end{cases}
\end{equation}
In our case, $a = -\frac{\gamma}{2v_g(\overline{M})}$, so as long as $\gamma>0$, $\lim_{x\rightarrow \infty} \bar{\mu}_g(x) = \lim_{x\rightarrow \infty} \bar{\mu}_s(x) = 0$ and thus MT lengths remain bounded. 

In the case where $\gamma = 0$, the model does not include a length-dependent catastrophe mechanism. In this case, $\lambdags(x) = \lambdags^0 + \gamma(x-L_0) = \lambdags^0$ and Eqs.~\eqref{eq:PDE_general} become:
\begin{equation}
\begin{aligned}
 \overline{\mu}_g(x) &= \frac{\nu}{v_g(\overline{M})} \exp\left(\left(\frac{\lambdasg}{v_s}- \frac{\lambdags^0}{v_g(\overline{M})}\right) x\right),\\
  \overline{\mu}_s(x) &= \frac{\nu}{v_s}     \exp\left(\left(\frac{\lambdasg}{v_s}- \frac{\lambdags^0}{v_g(\overline{M})}\right) x\right).\\
 \end{aligned}
\end{equation}
We are again interested in the behavior of these distributions in the limit that $x \rightarrow \infty$. To achieve bounded MT length distributions, we must have that:
\begin{equation}
\begin{aligned}
v_g(\overline{M})\lambdasg < \lambdags^0v_s.
\end{aligned}
\end{equation}

In Table~\ref{tab:nuc_params}, we note that $\lambdags^0 = 0.5 \text{ min}^{-1}$ and $v_s = 6 \mu\text{m}/\text{min}$ for our fruit fly neuron system. Using the parameterization procedure outlined in \cite{nelson2024minimal} with $L_* = 35 \mu$m and no length-dependent catastrophe, we find that $\lambdasg = 0.3592 \text{ min}^{-1}$. Given these parameter values, and to achieve bounded MT lengths, we must have that $v_g(\overline{M}) < 8.35 \mu\text{m}/\text{min}$, which we achieve through the restriction of growth speed  $v_g(\overline{M})$ in Eq.~\eqref{eq:vgofT} due to tubulin constraints.  In conclusion, the upper bound on $v_g(\overline{M})$ and the form of $v_g(\overline{M})$ guarantee that we obtain bounded model MT lengths in this application, even for the case that does not incorporate a length-dependent catastrophe mechanism. 

\bibliographystyle{unsrt}

\begin{thebibliography}{10}

\bibitem{kelliher2019microtubule}
Michael~T Kelliher, Harriet~AJ Saunders, and Jill Wildonger.
\newblock Microtubule control of functional architecture in neurons.
\newblock {\em Current Opinion in Neurobiology}, 57:39--45, 2019.

\bibitem{Rolls2021}
Melissa~M. Rolls, Pankajam Thyagarajan, and Chengye Feng.
\newblock Microtubule dynamics in healthy and injured neurons.
\newblock {\em Developmental Neurobiology}, 81:321--332, 4 2021.

\bibitem{mitchison1984dynamic}
Tim Mitchison and Marc Kirschner.
\newblock Dynamic instability of microtubule growth.
\newblock {\em Nature}, 312(5991):237--242, 1984.

\bibitem{teixido2012and}
Neus Teixid{\'o}-Travesa, Joan Roig, and Jens L{\"u}ders.
\newblock The where, when and how of microtubule nucleation--one ring to rule
  them all.
\newblock {\em Journal of Cell Science}, 125(19):4445--4456, 2012.

\bibitem{desai1997microtubule}
Arshad Desai and Timothy~J Mitchison.
\newblock Microtubule polymerization dynamics.
\newblock {\em Annual Review of Cell and Developmental Biology}, 13(1):83--117,
  1997.

\bibitem{weiner2021nucleate}
Alexis~T Weiner, Pankajam Thyagarajan, Yitao Shen, and Melissa~M Rolls.
\newblock To nucleate or not, that is the question in neurons.
\newblock {\em Neuroscience Letters}, 751:135806, 2021.

\bibitem{luders2021nucleating}
Jens L{\"u}ders.
\newblock Nucleating microtubules in neurons: Challenges and solutions.
\newblock {\em Developmental Neurobiology}, 81(3):273--283, 2021.

\bibitem{vinopal2025centrosomal}
Stanislav Vinopal and Frank Bradke.
\newblock Centrosomal and acentrosomal microtubule nucleation during neuronal
  development.
\newblock {\em Current Opinion in Neurobiology}, 92:103016, 2025.

\bibitem{gopalakrishnan2011first}
Manoj Gopalakrishnan and Bindu~S Govindan.
\newblock A first-passage-time theory for search and capture of chromosomes by
  microtubules in mitosis.
\newblock {\em Bulletin of Mathematical Biology}, 73:2483--2506, 2011.

\bibitem{mulder2012microtubules}
Bela~M Mulder.
\newblock Microtubules interacting with a boundary: Mean length and mean
  first-passage times.
\newblock {\em Physical Review E—Statistical, Nonlinear, and Soft Matter
  Physics}, 86(1):011902, 2012.

\bibitem{rubin1988}
Robert~J Rubin.
\newblock Mean lifetime of microtubules attached to nucleating sites.
\newblock {\em Proceedings of the National Academy of Sciences},
  85(2):446--448, 1988.

\bibitem{Dogterom1993}
Marileen Dogterom and Stanislas Leibler.
\newblock Physical aspects of the growth and regulation of microtubule
  structures microtubules (mts) are long, rigid polymers made of tubulin-a
  globular protein found in eukaryotic cells [1].
\newblock {\em Physical Review Letters}, 70:1347--1350, 1993.

\bibitem{zelinski2012dynamics}
Bj{\"o}rn Zelinski, Nina M{\"u}ller, and Jan Kierfeld.
\newblock Dynamics and length distribution of microtubules under force and
  confinement.
\newblock {\em Physical Review E—Statistical, Nonlinear, and Soft Matter
  Physics}, 86(4):041918, 2012.

\bibitem{jemseena2015effects}
V~Jemseena and Manoj Gopalakrishnan.
\newblock Effects of aging in catastrophe on the steady state and dynamics of a
  microtubule population.
\newblock {\em Physical Review E}, 91(5):052704, 2015.

\bibitem{Dogterom1995}
M~Dogterom, A~C Maggs, and S~Leibler.
\newblock Diffusion and formation of microtubule asters: physical processes
  versus biochemical regulation.
\newblock {\em Proceedings of the National Academy of Sciences}, 92:6683--6688,
  1995.

\bibitem{Deymier2005}
P.~A. Deymier, Y.~Yang, and J.~Hoying.
\newblock Effect of tubulin diffusion on polymerization of microtubules.
\newblock {\em Physical Review E - Statistical, Nonlinear, and Soft Matter
  Physics}, 72, 8 2005.

\bibitem{Margolin2006}
Gennady Margolin, Ivan~V. Gregoretti, Holly~V. Goodson, and Mark~S. Alber.
\newblock Analysis of a mesoscopic stochastic model of microtubule dynamic
  instability.
\newblock {\em Physical Review E - Statistical, Nonlinear, and Soft Matter
  Physics}, 74, 2006.

\bibitem{hinow2009continuous}
Peter Hinow, Vahid Rezania, and Jack~A Tuszy{\'n}ski.
\newblock Continuous model for microtubule dynamics with catastrophe, rescue,
  and nucleation processes.
\newblock {\em Physical Review E}, 80(3):031904, 2009.

\bibitem{Buxton2010}
Gavin~A. Buxton, Sandra~L. Siedlak, George Perry, and Mark~A. Smith.
\newblock Mathematical modeling of microtubule dynamics: Insights into
  physiology and disease.
\newblock {\em Progress in Neurobiology}, 92:478--483, 12 2010.

\bibitem{nelson2024minimal}
Anna~C Nelson, Melissa~M Rolls, Maria-Veronica Ciocanel, and Scott~A McKinley.
\newblock Minimal mechanisms of microtubule length regulation in living cells.
\newblock {\em Bulletin of Mathematical Biology}, 86(5):58, 2024.

\bibitem{bolterauer1999models}
H~Bolterauer, H-J Limbach, and JA~Tuszy{\'n}ski.
\newblock Models of assembly and disassembly of individual microtubules:
  stochastic and averaged equations.
\newblock {\em Journal of Biological Physics}, 25:1--22, 1999.

\bibitem{govindan2008length}
Bindu~S Govindan, Manoj Gopalakrishnan, and Debashish Chowdhury.
\newblock Length control of microtubules by depolymerizing motor proteins.
\newblock {\em Europhysics Letters}, 83(4):40006, 2008.

\bibitem{Mazilu2010}
I.~Mazilu, G.~Zamora, and J.~Gonzalez.
\newblock A stochastic model for microtubule length dynamics.
\newblock {\em Physica A: Statistical Mechanics and its Applications},
  389:419--427, 2 2010.

\bibitem{tischer2010providing}
Christian Tischer, Pieter~Rein Ten~Wolde, and Marileen Dogterom.
\newblock Providing positional information with active transport on dynamic
  microtubules.
\newblock {\em Biophysical Journal}, 99(3):726--735, 2010.

\bibitem{kuan2013biophysics}
Hui-Shun Kuan and MD~Betterton.
\newblock Biophysics of filament length regulation by molecular motors.
\newblock {\em Physical Biology}, 10(3):036004, 2013.

\bibitem{Rank2018}
Matthias Rank, Aniruddha Mitra, Louis Reese, Stefan Diez, and Erwin Frey.
\newblock Limited resources induce bistability in microtubule length
  regulation.
\newblock {\em Physical Review Letters}, 120, 4 2018.

\bibitem{honore2019growth}
St{\'e}phane Honor{\'e}, Florence Hubert, Magali Tournus, and Diana White.
\newblock A growth-fragmentation approach for modeling microtubule dynamic
  instability.
\newblock {\em Bulletin of Mathematical Biology}, 81:722--758, 2019.

\bibitem{gregoretti2006insights}
Ivan~V Gregoretti, Gennady Margolin, Mark~S Alber, and Holly~V Goodson.
\newblock Insights into cytoskeletal behavior from computational modeling of
  dynamic microtubules in a cell-like environment.
\newblock {\em Journal of Cell Science}, 119(22):4781--4788, 2006.

\bibitem{jain2021polymerization}
Kunalika Jain, Jashaswi Basu, Megha Roy, Jyoti Yadav, Shivprasad Patil, and
  Chaitanya~A Athale.
\newblock Polymerization kinetics of tubulin from mung seedlings modeled as a
  competition between nucleation and gtp-hydrolysis rates.
\newblock {\em Cytoskeleton}, 78(9):436--447, 2021.

\bibitem{Tao2016}
Juan Tao, Chengye Feng, and Melissa~M. Rolls.
\newblock The microtubule-severing protein fidgetin acts after dendrite injury
  to promote their degeneration.
\newblock {\em Journal of Cell Science}, 129:3274--3281, 2016.

\bibitem{chen2012axon}
Li~Chen, Michelle~C Stone, Juan Tao, and Melissa~M Rolls.
\newblock Axon injury and stress trigger a microtubule-based neuroprotective
  pathway.
\newblock {\em Proceedings of the National Academy of Sciences},
  109(29):11842--11847, 2012.

\bibitem{hertzler2020kinetochore}
James~I Hertzler, Samantha~I Simonovitch, Richard~M Albertson, Alexis~T Weiner,
  Derek~MR Nye, and Melissa~M Rolls.
\newblock Kinetochore proteins suppress neuronal microtubule dynamics and
  promote dendrite regeneration.
\newblock {\em Molecular biology of the cell}, 31(18):2125--2138, 2020.

\bibitem{Stone2010}
Michelle~C Stone, Michelle~M Nguyen, Juan Tao, Dana~L Allender, and Melissa~M
  Rolls.
\newblock Global up-regulation of microtubule dynamics and polarity reversal
  during regeneration of an axon from a dendrite.
\newblock {\em Molecular Biology of the Cell}, 21:767--777, 2010.

\bibitem{nguyen2014gamma}
Michelle~M Nguyen, Christie~J McCracken, ES~Milner, Daniel~J Goetschius,
  Alexis~T Weiner, Melissa~K Long, Nick~L Michael, Sean Munro, and Melissa~M
  Rolls.
\newblock $\gamma$-tubulin controls neuronal microtubule polarity independently
  of golgi outposts.
\newblock {\em Molecular Biology of the Cell}, 25(13):2039--2050, 2014.

\bibitem{weiner2020endosomal}
Alexis~T Weiner, Dylan~Y Seebold, Pedro Torres-Gutierrez, Christin Folker,
  Rachel~D Swope, Gregory~O Kothe, Jessica~G Stoltz, Madeleine~K Zalenski,
  Christopher Kozlowski, Dylan~J Barbera, et~al.
\newblock Endosomal wnt signaling proteins control microtubule nucleation in
  dendrites.
\newblock {\em PLOS Biology}, 18(3):e3000647, 2020.

\bibitem{kleele2014assay}
Tatjana Kleele, Petar Marinkovi{\'c}, Philip~R Williams, Sina Stern, Emily~E
  Weigand, Peter Engerer, Ronald Naumann, Jana Hartmann, Rosa~M Karl, Frank
  Bradke, et~al.
\newblock An assay to image neuronal microtubule dynamics in mice.
\newblock {\em Nature Communications}, 5(1):4827, 2014.

\bibitem{qu2019activity}
Xiaoyi Qu, Atul Kumar, Heike Blockus, Clarissa Waites, and Francesca Bartolini.
\newblock Activity-dependent nucleation of dynamic microtubules at presynaptic
  boutons controls neurotransmission.
\newblock {\em Current Biology}, 29(24):4231--4240, 2019.

\bibitem{Feng2019}
Chengye Feng, Pankajam Thyagarajan, Matthew Shorey, Dylan~Y. Seebold, Alexis~T.
  Weiner, Richard~M. Albertson, Kavitha~S. Rao, Alvaro Sagasti, Daniel~J.
  Goetschius, and Melissa~M. Rolls.
\newblock Patronin-mediated minus end growth is required for dendritic
  microtubule polarity.
\newblock {\em Journal of Cell Biology}, 218:2309--2328, 2019.

\bibitem{kshirsagar2024resolving}
Sudhir Kshirsagar, Md~Ariful Islam, Arubala~P Reddy, and P~Hemachandra Reddy.
\newblock Resolving the current controversy of use and reuse of housekeeping
  proteins in ageing research: Focus on saving people’s tax dollars.
\newblock {\em Ageing Research Reviews}, page 102437, 2024.

\bibitem{walker1988dynamic}
RA~Walker, ET~O'brien, NK~Pryer, MF~Soboeiro, WA~Voter, HP~Erickson, and
  Edward~D Salmon.
\newblock Dynamic instability of individual microtubules analyzed by video
  light microscopy: rate constants and transition frequencies.
\newblock {\em The Journal of Cell Biology}, 107(4):1437--1448, 1988.

\bibitem{white2017exploring}
Diana White, St{\'e}phane Honor{\'e}, and Florence Hubert.
\newblock Exploring the effect of end-binding proteins and microtubule
  targeting chemotherapy drugs on microtubule dynamic instability.
\newblock {\em Journal of Theoretical Biology}, 429:18--34, 2017.

\bibitem{yau2014microtubule}
Kah~Wai Yau, Sam~FB van Beuningen, In{\^e}s Cunha-Ferreira, Bas~MC Cloin,
  Eljo~Y van Battum, Lena Will, Philipp Sch{\"a}tzle, Roderick~P Tas, Jaap van
  Krugten, Eugene~A Katrukha, et~al.
\newblock Microtubule minus-end binding protein camsap2 controls axon
  specification and dendrite development.
\newblock {\em Neuron}, 82(5):1058--1073, 2014.

\bibitem{bressloff2019search}
Paul~C Bressloff and Hyunjoong Kim.
\newblock Search-and-capture model of cytoneme-mediated morphogen gradient
  formation.
\newblock {\em Physical Review E — Statistical, Nonlinear, and Soft Matter
  Physics}, 99(5):052401, 2019.

\bibitem{tindemans2010microtubule}
Simon~H Tindemans and Bela~M Mulder.
\newblock Microtubule length distributions in the presence of protein-induced
  severing.
\newblock {\em Physical Review E—Statistical, Nonlinear, and Soft Matter
  Physics}, 81(3):031910, 2010.

\bibitem{gardner2011depolymerizing}
Melissa~K Gardner, Marija Zanic, Christopher Gell, Volker Bormuth, and Jonathon
  Howard.
\newblock Depolymerizing kinesins kip3 and mcak shape cellular microtubule
  architecture by differential control of catastrophe.
\newblock {\em Cell}, 147(5):1092--1103, 2011.

\bibitem{gardner2013microtubule}
Melissa~K Gardner, Marija Zanic, and Jonathon Howard.
\newblock Microtubule catastrophe and rescue.
\newblock {\em Current Opinion in Cell Biology}, 25(1):14--22, 2013.

\bibitem{cassimeris1986dynamics}
Lynne~U Cassimeris, Patricia Wadsworth, and ED~Salmon.
\newblock Dynamics of microtubule depolymerization in monocytes.
\newblock {\em The Journal of Cell Biology}, 102(6):2023--2032, 1986.

\bibitem{schulze1986microtubule}
Eric Schulze and Marc Kirschner.
\newblock Microtubule dynamics in interphase cells.
\newblock {\em The Journal of Cell Biology}, 102(3):1020--1031, 1986.

\bibitem{kuo2019spastin}
Yin-Wei Kuo, Olivier Trottier, Mohammed Mahamdeh, and Jonathon Howard.
\newblock Spastin is a dual-function enzyme that severs microtubules and
  promotes their regrowth to increase the number and mass of microtubules.
\newblock {\em Proceedings of the National Academy of Sciences},
  116(12):5533--5541, 2019.

\bibitem{verde1990regulation}
Fulvia Verde, Jean-Claude Labb{\'e}, Marcel Dor{\'e}e, and Eric Karsenti.
\newblock Regulation of microtubule dynamics by cdc2 protein kinase in
  cell-free extracts of xenopus eggs.
\newblock {\em Nature}, 343(6255):233--238, 1990.

\bibitem{gliksman1992okadaic}
Neal~R Gliksman, Stephen~F Parsons, and ED~Salmon.
\newblock Okadaic acid induces interphase to mitotic-like microtubule dynamic
  instability by inactivating rescue.
\newblock {\em The Journal of Cell Biology}, 119(5):1271--1276, 1992.

\bibitem{howard2007microtubule}
Jonathon Howard and Anthony~A Hyman.
\newblock Microtubule polymerases and depolymerases.
\newblock {\em Current Opinion in Cell Biology}, 19(1):31--35, 2007.

\bibitem{gillespie2001approximate}
Daniel~T Gillespie.
\newblock Approximate accelerated stochastic simulation of chemically reacting
  systems.
\newblock {\em The Journal of Chemical Physics}, 115(4):1716--1733, 2001.

\bibitem{gillespie2007stochastic}
Daniel~T Gillespie et~al.
\newblock Stochastic simulation of chemical kinetics.
\newblock {\em Annual Review of Physical Chemistry}, 58(1):35--55, 2007.

\end{thebibliography}

\end{document}